\documentclass[preprint,12pt]{elsarticle} 

\usepackage{lineno}
\modulolinenumbers[5]
\usepackage{color}
\usepackage[usenames,dvipsnames]{xcolor}
\usepackage{fullpage}
\usepackage{amsmath}
\usepackage{amssymb}
\usepackage{mathrsfs}
\usepackage{newfloat}
\usepackage{graphicx}
\usepackage{subfig}
\usepackage{multirow}
\usepackage{stackengine}
\usepackage{caption}
\usepackage{tikz}
\usepackage{multicol}
\usetikzlibrary{arrows,positioning,calc}
\usepackage{pstool}
\usepackage{textcomp}
\usepackage[hidelinks]{hyperref}
\hypersetup{colorlinks,bookmarksopen,linkcolor=blue,citecolor=blue,urlcolor=blue}

\definecolor{myblue}{RGB}{0,115,189}
\definecolor{mygreen}{rgb}{0.19,0.61,0.21}

\newcommand{\bs}[1]{{\boldsymbol{#1}}}
\newcommand{\tr}[1]{\mathrm{tr\,{#1}}}

\def\Aop{\operatornamewithlimits{\mathchoice{\vcenter{\hbox{\huge \sf{A}}}}{\vcenter{\hbox{\Large A}}}{\mathrm{A}}{\mathrm{A}}}}
\DeclareFloatingEnvironment{algorithm}

\journal{J. Mech. Phys. Solids}

\biboptions{authoryear}

\begin{document}

\begin{frontmatter}

\title{Micromorphic Computational Homogenization for Mechanical Metamaterials with Patterning Fluctuation Fields\tnoteref{titlefoot}}

\author[TUe]{O.~Roko\v{s}} 
\ead{O.Rokos@tue.nl}

\author[TUe]{M.M.~Ameen}
\ead{M.Ameen@tue.nl}

\author[TUe]{R.H.J.~Peerlings}
\ead{R.H.J.Peerlings@tue.nl}

\author[TUe]{M.G.D.~Geers\corref{correspondingauthor}}
\ead{M.G.D.Geers@tue.nl}

\address[TUe]{Mechanics of Materials, Department of Mechanical Engineering, Eindhoven University of Technology, P.O.~Box~513, 5600~MB~Eindhoven, The~Netherlands}
\cortext[correspondingauthor]{Corresponding author.}

\tnotetext[titlefoot]{The post-print version of this article is published in \emph{J. Mech. Phys. Solids}, \href{https://www.sciencedirect.com/science/article/pii/S0022509618306148}{10.1016/j.jmps.2018.08.019}.}

\begin{abstract}
This paper presents a homogenization framework for elastomeric metamaterials exhibiting long-range correlated fluctuation fields. Based on full-scale numerical simulations on a class of such materials, an ansatz is proposed that allows to decompose the kinematics into three parts, i.e. a smooth mean displacement field, a long-range correlated fluctuating field, and a local microfluctuation part. With this decomposition, a homogenized solution is defined by ensemble averaging the solutions obtained from a family of translated microstructural realizations. Minimizing the resulting homogenized energy, a micromorphic continuum emerges in terms of the average displacement and the amplitude of the patterning long-range microstructural fluctuation fields. Since full integration of the ensemble averaged global energy (and hence also the corresponding Euler--Lagrange equations) is computationally prohibitive, a more efficient approximative computational framework is developed. The framework relies on local energy density approximations in the neighbourhood of the considered Gauss integration points, while taking into account the smoothness properties of the effective fields and periodicity of the microfluctuation pattern. Finally, the implementation of the proposed methodology is briefly outlined and its performance is demonstrated by comparing its predictions against full scale simulations of a representative example. 

\end{abstract}

\begin{keyword}
Mechanical metamaterials \sep computational homogenization \sep micromorphic continuum \sep  non-linear homogenization 
\end{keyword}

\end{frontmatter}

%
%
\section{Introduction}
\label{introduction}
Mechanical metamaterials exhibiting exotic mechanical properties, such as auxetic behaviour, serve dedicated applications, most notably in soft robotics~\citep{Whitesides:2018,Mirzaali:2018,Zhang:2018,Mark:2016,Jiang:2016,Yang:2015}. The development of advanced manufacturing techniques, like~3D printing, has enabled a significant increase of the design space for making materials with certain specific architected microstructures~\citep{Ren:2018,Truby:2016,Wang:2015}. In this contribution, we in particular concentrate on elastomeric materials with periodic microstructures consisting of a regular grid of circular holes, which undergo a \textit{pattern transformation} triggered by externally applied compression upon reaching a critical compressive strain~\citep{Bertoldi:2010a}. The pattern-transformed material has been shown to reveal a significantly different effective behaviour compared to its initial untransformed state~\citep{Bertoldi2008d}. 

\begin{figure}[ht]
	\centering
	\subfloat[displacement field]{\includegraphics[width=0.2\textwidth]{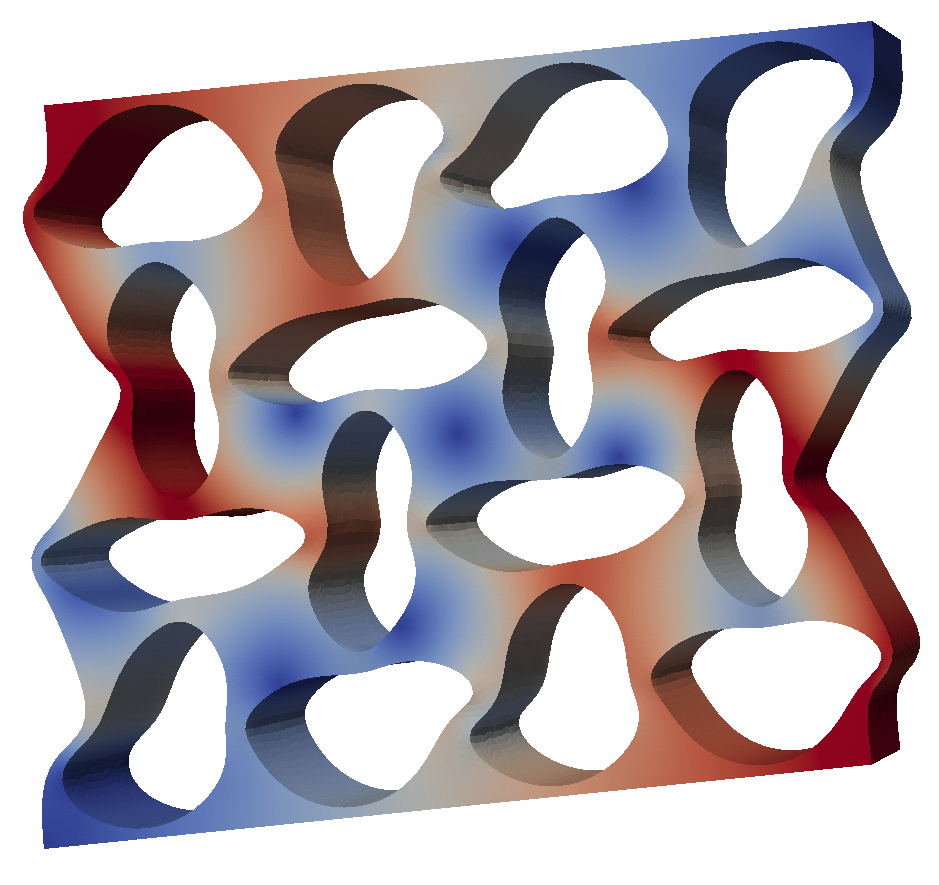}\label{introduction:fig1a}}\hspace{2em}
    \subfloat[mean displacement field]{\includegraphics[width=0.2\textwidth]{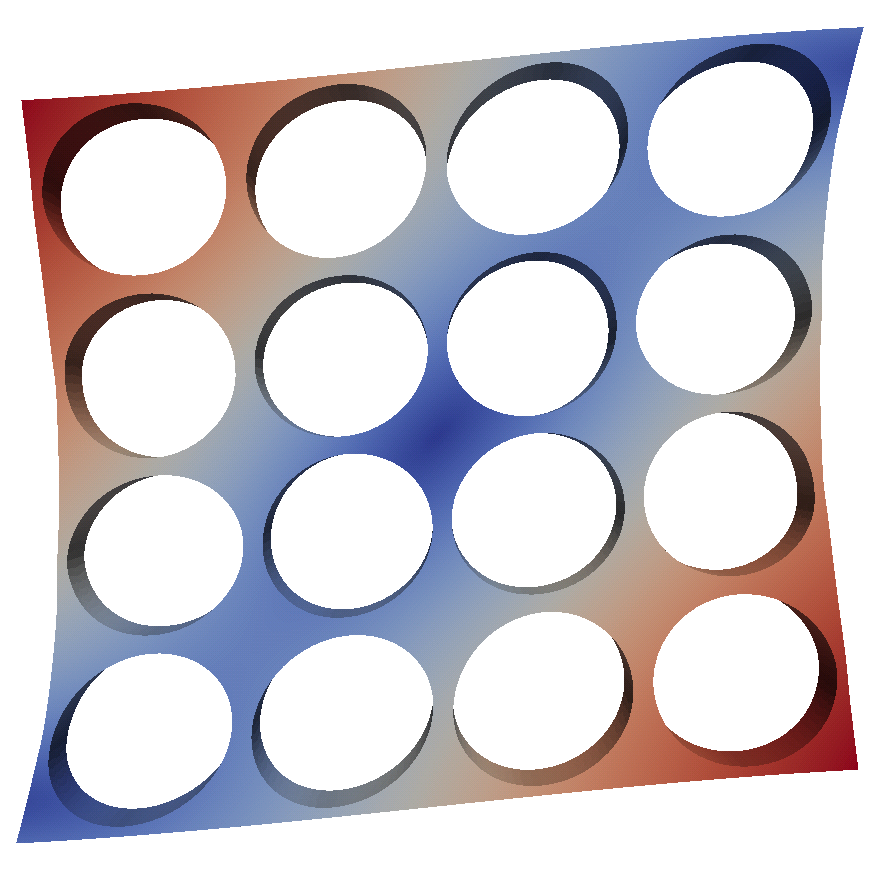}\label{introduction:fig1b}}\hspace{2em}
	\subfloat[long-range correlated fluctuation field]{\includegraphics[width=0.2\textwidth]{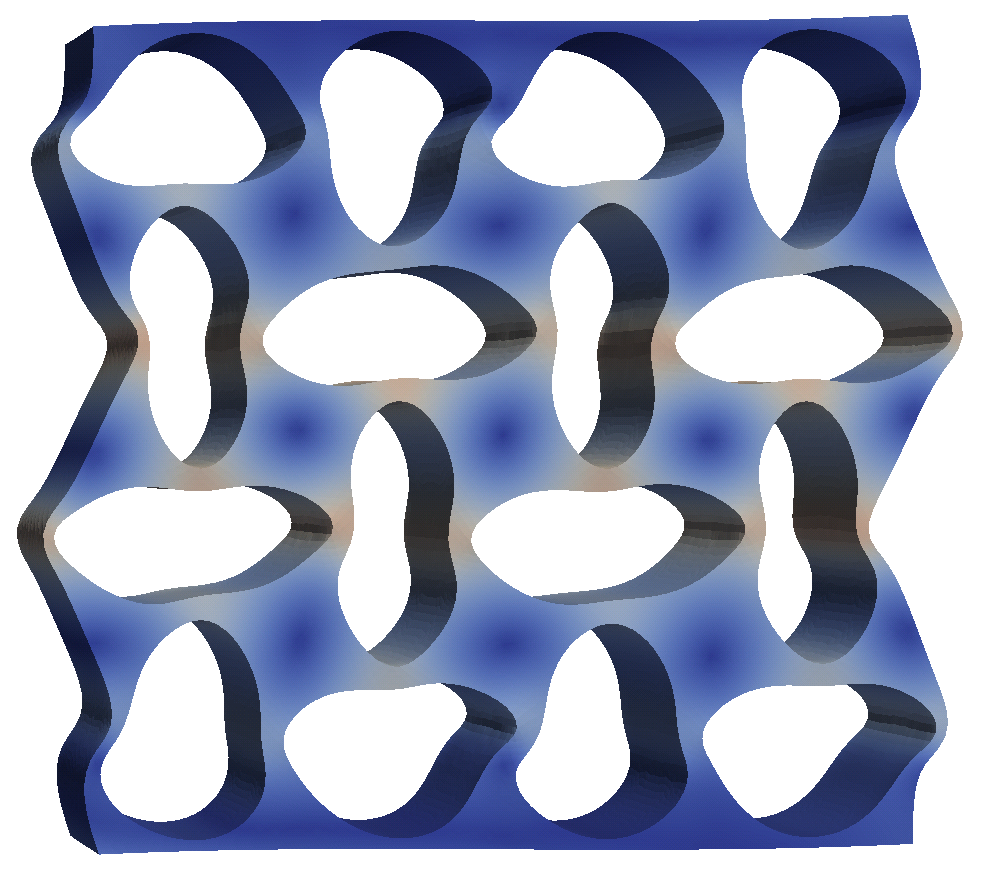}\label{introduction:fig1c}}\hspace{0.75em}
	\stackunder{\includegraphics[height=3.25cm]{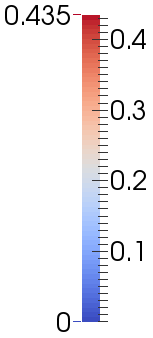}}{\small $\|\vec{u}\|_2/\ell$}
	\caption{Kinematical decomposition of (a)~the general displacement field into its (b)~effective (mean field) part and (c)~long-range correlated fluctuating part for a square specimen of size~$L$, having a scale ratio~$L/\ell = 4$. The local microfluctuating uncorrelated part is condensed out (not shown). The colour indicates the (pointwise) Euclidean norm of the displacement field~$\vec{u}$ relative to the primitive cell size~$\ell$.}
	\label{introduction:fig1}
\end{figure}

The overall effective response of this class of materials depends on the number of holes (which may be intractable) as well as their relative (unknown) position with respect to the specimen's geometry. This is a challenging problem because, depending on the loading and boundary conditions applied to the specimen, the patterning may occur or it may be restricted. Adjacent cells kinematically communicate in this respect and the effective response is intrinsically non-local. A typical example is shown in Fig.~\ref{introduction:fig1a}, where an elastomeric sheet of size~$L \times L$ with circular holes subjected to combined compression and shear is depicted. The resulting deformation exhibits a transformed pattern, a shear band, and stiff boundary layers. The boundary layers are present due to the kinematic constraints close to the physical boundaries (the two horizontal edges), where the regular buckling pattern cannot fully develop. The non-transformed material in such constrained boundaries has a higher stiffness, leading directly to stiff boundary layers. These non-uniformities in the deformation pattern give rise to a size effect in the overall mechanical behaviour of the considered system, i.e. the overall response depends on the scale ratio. The scale ratio is, in this particular case, defined as the total height or width of the entire specimen~$L$ with respect to the size of one primitive cell~$\ell$ (containing only one hole), i.e. as~$L/\ell$. In order to obtain the effective macroscopic response, use can be made of the periodicity of the microstructure. For the example shown, a microstructural periodic cell consisting of $2 \times 2$ holes can be evaluated to obtain the homogenized response, as shown experimentally by~\cite{Bertoldi2008d}, and computationally by~\cite{Ameen2018} using ensemble averaging. 

A fruitful starting point for computing the effective response of a heterogeneous material is provided by computational homogenization. Conventional first order computational homogenization~\citep{Kouznetsova:2001,Miehe:2002,Matsui:2004}, can provide accurate results for elastomeric metamaterials only for very large scale ratios~\citep{Ameen2018}. Here, a microscale boundary value problem is solved at every macroscopic point to retrieve the macroscopic constitutive response, which is used in a concurrent macroscopic simulation. For smaller scale ratios, however, the long-range fluctuation field, resulting from the kinematic interaction between neighbouring microstructural periodic cells, needs to be captured by the homogenization framework. Various higher-order homogenization frameworks have been proposed in the literature, which help to capture the non-local behaviour. Second-order computational homogenization~\citep{Kouznetsova:2004}, for instance, incorporates the gradient of the macroscopic deformation gradient tensor into the kinematic macro-micro scale transition. It entails, however, additional complexity through the solution of a higher-order equilibrium equation at the macroscale, for which the length scale equals the size of the Representative Volume Element~(RVE), see e.g~\cite{Kouznetsova:2004size} for further details.

Another class of extended continua has been proposed via generalized micromorphic theories by~\cite{Eringen:1968} and~\cite{Forest:2009}. Here, additional kinematic fields are introduced in order to incorporate underlying microfluctuation fields which are not captured by standard kinematic variables. The Cosserat continuum~\citep{Cosserat:1909} is a well-known example, where the additional kinematic field corresponds to local rotations. \cite{Hutter:2017} proposed a methodology for homogenization towards a fully micromorphic continuum and derived the relevant generalized stresses. \cite{Biswas:2017} proposed a computational homogenization scheme for matrix-inclusion composites, in which an additional kinematic field characterizing the average strain in the inclusions is introduced, also resulting in a micromorphic continuum. It was shown that this framework is capable of capturing the homogenized response for such materials without a clear separation of length scales. Other relevant contributions can be found in \cite{Janicke:2012} or~\cite{Forest:2002}, and in the references therein.

The present paper proposes a novel computational homogenization framework for materials exhibiting long-range correlated fluctuation fields. Unlike the above-mentioned references, the adopted approach directly incorporates the dominant underlying microstructural kinematical fluctuations through a characteristic fluctuation mode. This mode uniquely relates to the morphology of the microstructure, and is not necessarily limited to rotation or higher-order deformation only. Any relevant kinematical interaction between adjacent periodic cells can be captured. To this end, first the displacement field is decomposed into three sub-fields: a mean effective field, a long-range spatially correlated field, and a local microfluctuation field, cf. Figs.~\ref{introduction:fig1b} and~\ref{introduction:fig1c}. In the next step, an entire family of translated microstructures relative to the specimen geometry is considered, the average of which defines the effective behaviour. The entire problem is then formulated variationally, which upon minimization yields the associated Euler--Lagrange equations of the effective micromorphic continuum in terms of the mean displacement and the amplitude of the correlated part of the fluctuation, along with the associated boundary conditions. To avoid having to solve for the microfluctuation field in the entire domain, which effectively makes the solution computationally as expensive as full-scale simulations, two approximations are considered. The first one neglects the uncorrelated microfluctuation field completely. This works well in terms of kinematic quantities, but it turns out that the conjugate quantities are very inaccurate. A second approach is therefore elaborated, in which an approximate microfluctuation field is computed only locally, inside a microstructural periodic cell. To render the computation more efficient, approximations of the local average energy density in the neighbourhood of the considered macroscopic Gauss integration points are further considered (with respect to the microstructural translations), while taking into account the smoothness properties of the effective fields and periodicity of the microfluctuation field. The performance of the proposed computational homogenization framework is evaluated using a representative example, by comparing against the Direct Numerical Simulations~(DNS) published by~\cite{Ameen2018}. The effective kinematic quantities are thereby captured with adequate accuracy. Moreover, the homogenized nominal stresses now also closely predict the size effect exhibited by the DNS solution, achieving a maximum relative error of less than~$10\,\%$ for the smallest scale ratio considered.

The paper is divided into five sections as follows. The next section, Section~\ref{problem}, describes the full-scale problem considered, its corresponding DNS solution, and the methodology used to define the mean (reference) solution using ensemble averaging. Section~\ref{micromorphic} then presents the decomposition of the kinematics into the three sub-fields and provides the derivation of the full generalized homogenization framework, which leads to the emergence of an effective micromorphic continuum. Because the resulting problem is still computationally expensive, Section~\ref{simplified} introduces a simplified approach in which the microfluctuation field is neglected and considered as a truncation error. The results obtained with this approach suggest the need to incorporate the fast fluctuating field, since it contributes significantly in terms of energy. Section~\ref{fe2} therefore elaborates on a computationally integrated formulation, in which the microfluctuation field is computed only locally, inside periodic cells associated with individual Gauss integration points. This formulation is dual to the computational homogenization solution of the micromorphic continuum. The performance of the proposed methodology is compared against full-scale simulations in Section~\ref{results}, and the paper closes with a summary and conclusions in Section~\ref{summary}.

Throughout the paper, the following notational conventions will be used
\begin{multicols}{2}
\begin{itemize}[\textbf{-}]
\itemsep0em 
\item scalars~$ a $,
\item vectors~$ \vec{a} $,
\item second-order tensors~$ \bs{A} $,
\item fourth-order tensors~$ ^4\bs{C} $,
\item matrices~$ \bs{\mathsf{A}} $ and column matrices~$ \underline{a} $,
\item $ \vec{a} \cdot \vec{b} = a_i b_i $,
\item Euclidean norm~$ \| \vec{a} \|_2 = \sqrt{ \vec{a} \cdot \vec{a} }$,
\item $ \bs{A} \cdot \vec{b} =  A_{ij} b_j\vec{e}_i $,
\item $\bs{A}\cdot\bs{B} = A_{ik}B_{kj}\vec{e}_i\vec{e}_j $,
\item $\bs{A}:\bs{B} = A_{ij}B_{ji}$,
\item conjugate~$ \bs{A}^\mathsf{T}$, $ A_{ij}^\mathsf{T} = A_{ji} $,
\item gradient operator~$ \displaystyle \vec{\nabla} \vec{a} = \frac{\partial a_j}{\partial X_i} \vec{e}_i \vec{e}_j $,
\item divergence operator~$ \displaystyle \vec{\nabla} \cdot \vec{a} = \frac{\partial a_i}{\partial X_i} $,
\item derivatives of scalar functions with respect to second-order tensors \\
$\displaystyle \delta\Psi(\bs{F};\delta\bs{F}) = \left.\frac{\mathrm{d}}{\mathrm{d}h}\Psi(\bs{F}+h\delta\bs{F})\right|_{h=0} = \frac{\partial\Psi(\bs{F})}{\partial\bs{F}}:\delta\bs{F} $.
\end{itemize}
\end{multicols}
%
%
\section{Problem Statement}
\label{problem}
A representative example consisting of an elastomeric metamaterial exhibiting a patterned deformation under compressive loading is described in Section~\ref{example}, which is used as the reference problem. Although the developments made throughout this paper are only demonstrated on this simple example, the entire methodology proposed is equally applicable to more complex cases. The full-scale problem for a low scale separation regime, along with the DNS solutions are briefly discussed in Section~\ref{full_problem}. Ensemble averaging and the adopted definition for the effective reference solution are presented in Section~\ref{full_ensemble}.
%
%
\subsection{Reference Problem Definition}
\label{example}
\begin{figure}
	\centering
	\subfloat[Sketch of the geometry]{\def\svgwidth{0.3\textwidth}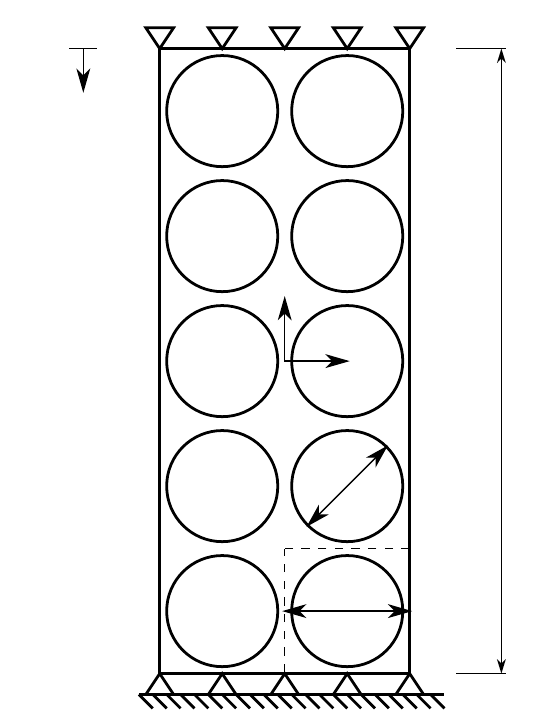\label{problem:fig1a}}\hspace{0.5em}
    \subfloat[Deformed mesh]{\includegraphics[height=6.17cm]{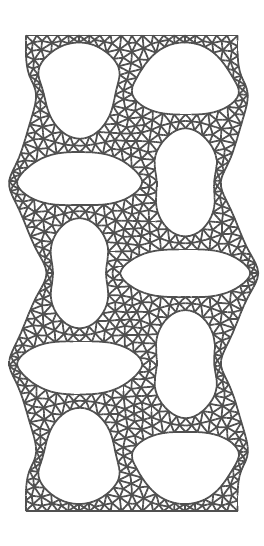}\label{problem:fig1b}}\hspace{0.5em}
	\subfloat[Effective stress--strain diagram]{\includegraphics[scale=1]{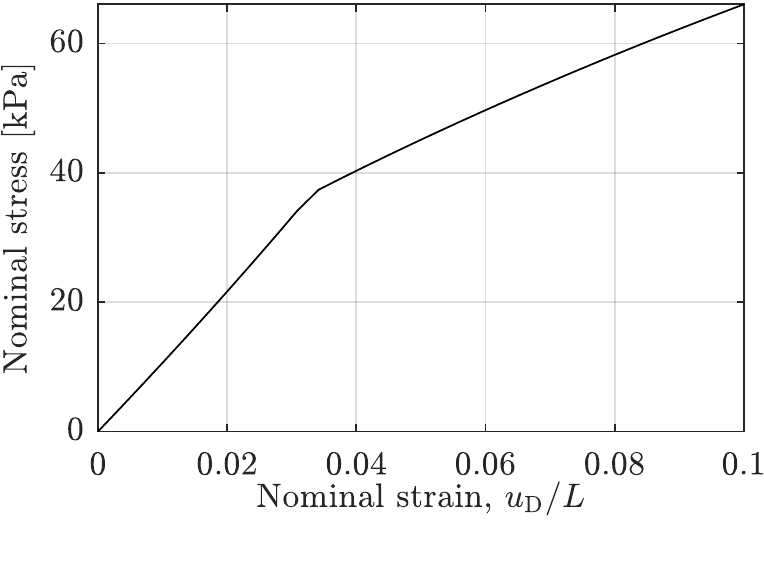}\label{problem:fig1c}}
	\caption{A sketch of a semi-infinite specimen~$\Omega_\mathcal{S} = (-\infty,\infty) \times [-L/2,L/2]$ subjected to compression, and modelled as a finite domain~$\Omega = [-\ell,\ell] \times [-L/2,L/2]$ with Periodic Boundary Conditions~(PBC). (a)~Geometry, (b)~deformed configuration corresponding to an overall strain of~$7.5\,\%$, and~(c) compressive nominal stress as a function of compressive nominal strain~$u_\mathrm{D}/L$.}
	\label{problem:fig1}
\end{figure}

The reference problem geometry is sketched in Fig.~\ref{problem:fig1a}. A semi-infinite specimen~$\Omega_\mathcal{S} = (-\infty,\infty) \times [-L/2,L/2]$ is considered. Since the specimen geometry is infinitely wide (along the horizontal axis~$\vec{e}_1$) and the applied deformation is uniform in this direction, a model domain~$\Omega = [-\ell,\ell] \times [-L/2,L/2]$, with Periodic Boundary Conditions~(PBC) along the two vertical edges (i.e. along~$AD$ and~$BC$), is considered. This is justified experimentally and computationally, since the deformation field typically exhibits a~$2\ell \times 2\ell$ repetitive pattern, see e.g.~\cite{Bertoldi2008c} or~\cite{Ameen2018}. The specimen is subjected to uniaxial compression by prescribing vertical displacements along the two horizontal edges~$\Gamma_\mathrm{D} = AB \cup CD \subset \partial\Omega$, as~$u_\mathrm{D}$ (on the top edge~$AB$) and zero~(on the bottom edge~$CD$), while constraining the horizontal displacements on both these edges. The considered domain consists of a number of square single-hole primitive cells, the size of which is the characteristic microstructural length~$\ell$. The macroscopic length, on the other hand, is specified by the height of the specimen~$L$, giving rise to a scale ratio~$L/\ell$. In what follows, only positive integer scale ratios larger than~$3$ will be considered, i.e.~$L/\ell \in \mathbb{N}_{>3}$.

The constitutive behaviour of the elastomeric base material is assumed to be hyperelastic, described by the strain energy density
\begin{equation}
\psi( \bs{F} (\vec{X}) ) = c_1 (I_1 - 3) + c_2 (I_1 - 3)^2 - 2 c_1 \log{J} + \frac{1}{2} K (J-1)^2,
\label{proglem:eq1}  
\end{equation}
whereas the individual holes are captured through the topology of the microstructure. In Eq.~\eqref{proglem:eq1}, $\vec{X} \in \Omega$, $\vec{X} = X_1\vec{e}_1 + X_2\vec{e}_2$ is the position vector in the reference configuration, $ \bs{F} = \bs{I} + ( \vec{\nabla} \vec{u} )^\mathsf{T} $ is the deformation gradient tensor, $\vec{\nabla}$ indicates the gradient with respect to the reference configuration, $\vec{u}(\vec{X})$ is the displacement field, $ J = \det{\bs{F}}$, $\bs{I}$ is the second-order identity tensor, and~$I_1 = \tr{\bs{C}}$ and~$I_2 = \frac{1}{2} \left[(\tr{\bs{C}})^2 - \tr{(\bs{C}^2)} \right]$ are the invariants of the right Cauchy-Green deformation tensor~$ \bs{C} = \bs{F}^\mathsf{T} \cdot \bs{F} $. 

The constitutive and geometric parameters adopted throughout this contribution are listed in Tab.~\ref{problem:tab1}, based on the experimental characterization by~\cite{Bertoldi2008d}. Note that the same parameters have been used for generating the full-scale DNS solutions reported in~\cite{Ameen2018}.
\begin{table}
\centering
\caption{Constitutive and geometrical parameters.}
\label{problem:tab1}
\begin{tabular}{c|ccc|cc}
\multirow{2}{*}{Parameter} & $ c_1 $ & $ c_2 $ & $ K $ & $ \ell $ & $ d $ \\
& [MPa] & [MPa] & [MPa] & [mm] & [mm] \\\hline
Value & $0.55$ & $0.3$ & $55$ & $9.97$ & $8.67$
\end{tabular}
\end{table}
%
%
\subsection{Full-scale Problem}
\label{full_problem}
For the above hyperelastic problem, the solution (displacement field) $\vec{u}(\vec{X})$ is obtained by minimizing the total potential energy of the entire system, i.e.
\begin{equation}
\begin{aligned}
\vec{u}(\vec{X}) &\in \underset{\vec{\widehat{u}}(\vec{X}) \in \mathscr{U}(\Omega)}{\mbox{arg min}} \ \mathcal{E}(\vec{\widehat{u}}(\vec{X})), \\
\mathcal{E}(\vec{\widehat{u}}) &= \int_{\Omega}\widehat{\Psi}(\vec{X}, \bs{F} (\vec{\widehat{u}}(\vec{X})) )\,\mathrm{d}\vec{X}, \quad \widehat{\Psi}(\vec{X}, \bs{F}) = \chi(\vec{X})\psi(\bs{F}),
\end{aligned}
\label{problem:eq2}  
\end{equation}
where~$\chi$ is the indicator function of the base material, i.e.~$\chi = 0$ inside holes and~$\chi = 1$ outside (implying also that all internal holes remain stress-free), and~$\mathscr{U}(\Omega)$ denotes the space of kinematically admissible displacement fields over the domain~$\Omega$ (i.e.~the vector space of all vector functions that satisfy the boundary conditions applied on the two horizontal edges, while being periodic in the horizontal direction with the period~$2\ell$). In anticipation of ensemble averaging in Section~\ref{full_ensemble}, where knowledge of the displacement field~$\vec{u}(\vec{X})$ at each position~$\vec{X} \in \Omega$ (including holes) is required, the entire rectangle~$\Omega = [-\ell,\ell]\times[-L/2,L/2]$ is considered as the domain of~$\vec{u}$, i.e.~$\vec{u}(\vec{X})$, $\vec{X} \in \Omega$. To this end, displacements and strains inside holes are interpolated using an ultra-soft linear-elastic material, whereas all stress components are set to zero there. As in Eq.~\eqref{problem:eq2}, throughout this manuscript admissible test functions are denoted with hats ($\widehat{\bullet}$), whereas minimizers are free of hats. The inclusion sign~$\in$ in Eq.~\eqref{problem:eq2} emphasizes the non-uniqueness resulting from non-convexity of the total potential energy~$\mathcal{E}$, which allows for multiple solutions, instabilities, and buckling.

The typical behaviour of the solutions for the considered problem is shown in Fig.~\ref{problem:fig1}. Here, an approximately bi-linear nominal stress--strain response is observed in Fig.~\ref{problem:fig1c}, the individual linear regimes being separated by a critical strain of about~$3\,\%$. At this strain, the microstructure buckles and nucleates a regular deformation pattern, as shown in Fig.~\ref{problem:fig1b}, which results in a significant change in nominal stiffness. Due to kinematic constraints close to the physical boundaries, the regular buckling pattern cannot fully develop. This implies that such constrained boundaries directly lead to more stiff boundary layers, as the non-transformed material has a higher stiffness. Clearly, considering more complex loading scenarios would entail more complicated mechanisms, including boundary layers, localization bands (recall Fig.~\ref{introduction:fig1}) and percolation paths. The presence of such non-uniformities in the deformation pattern furthermore gives rise to a size effect in the overall mechanical behaviour of the system, i.e. a dependence of this overall response on the scale ratio~$L/\ell$---particularly for ratios~$L/\ell < 10$. Because most of these phenomena result from the non-linearities and kinematic interactions between individual neighbouring periodic cells, the considered problem is challenging from a homogenization point of view.
\begin{figure}
	\flushleft\hspace{3.25em}
	\includegraphics[scale=1]{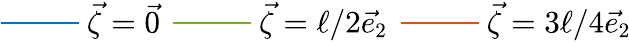}\vspace{-1.5em}\\
	\centering
    \subfloat[vertical displ.]{\includegraphics[scale=1]{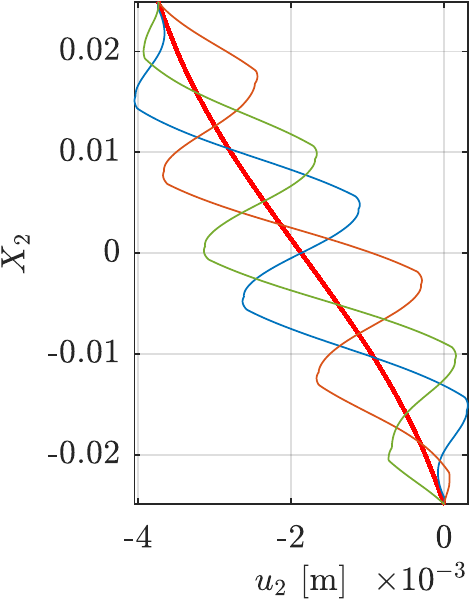}\label{problem:fig2a}}\hspace{0.5em}
	\subfloat[std. dev.]{\includegraphics[scale=1]{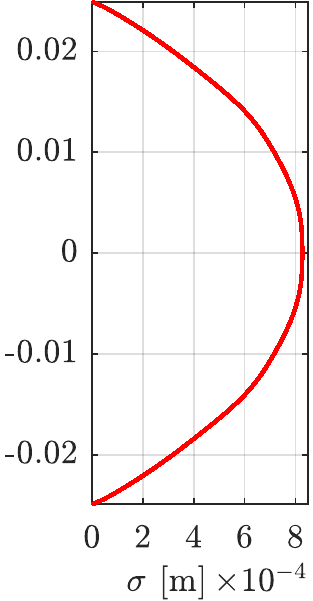}\label{problem:fig2b}}\hspace{0.5em}
	\subfloat[DNS mode]{\includegraphics[scale=1]{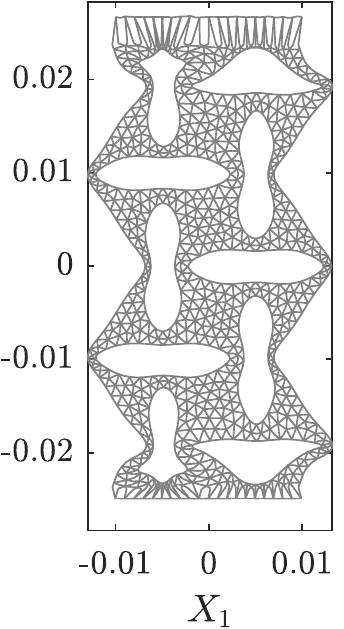}\label{problem:fig2c}}\hspace{0.5em}
	\subfloat[analytical mode]{\includegraphics[scale=1]{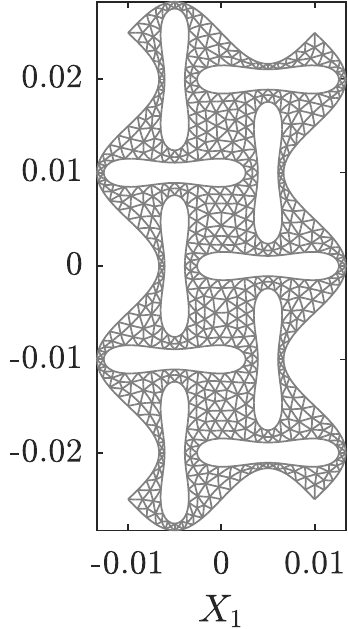}\label{problem:fig2d}}\hspace{0.5em}
	\caption{Components of displacement fields obtained for various translations of the microstructure at~$X_1 = -\ell/2$, $X_2 \in [-L/2,L/2]$: (a)~vertical component, (b)~standard deviation of~$\vec{u}-\vec{\overline{u}}$, denoted~$\sigma(\vec{X})$. The effective solution~$\vec{\overline{u}}$ is shown in thick red, whereas solutions for individual translations are shown in various colours. (c)~Typical DNS buckling mode~$\vec{\varphi}_1$, estimated as~$[\vec{u}(\vec{\zeta},\vec{X})-\vec{\overline{u}}(\vec{X})]/\sigma(\vec{X})$, and~(d)~analytical approximation of the mode~$\vec{\varphi}_1$ through Eq.~\eqref{problem:eq6}.}
	\label{problem:fig2}
\end{figure}
%
%
\subsection{Ensemble Averaging}
\label{full_ensemble}
One might not be, in general, able to control the positioning of the boundary relative to the microstructure---and the other way around. Hence, all possible realizations are considered and the reference mean solution is defined as their ensemble average. In order to establish an effective (homogenized) solution, from which all the fluctuations due to the microstructure are eliminated, a family of minimization problems with translated microstructures similar to Eq.~\eqref{problem:eq2} are therefore considered. The translation of the microstructure relative to the specimen's boundaries is denoted by~$\vec{\zeta} \in Q$, $\vec{\zeta} = \zeta_1\vec{e}_1+\zeta_2\vec{e}_2$, where~$Q = [-\ell,\ell]\times[-\ell,\ell]$ is the corresponding $2 \times 2$ \textit{periodic cell}. The periodic cell should not be confused with the \textit{primitive cell} (defined as~$[0,\ell]\times[0,\ell]$), which reflects the geometrical periodicity in the undeformed state. The periodic cell has the period of the fluctuating pattern, which is important for the homogenization, and which relies on experimental and numerical evidence. In general cases, Bloch analysis~\citep[cf. e.g.][]{Geymonat:1993,Singamaneni:2009a} can be used to estimate the wavelength of the pattern. The minimization problems associated with the individual translated microstructures are specified as
\begin{equation}
\begin{aligned}
\vec{u}(\vec{X},\vec{\zeta}) &\in \underset{\vec{\widehat{u}}(\vec{\zeta},\vec{X}) \in \mathscr{U}(\Omega)}{\mbox{arg min}} \ \mathcal{E}(\vec{\zeta},\vec{\widehat{u}}(\vec{X},\vec{\zeta})), \\
\mathcal{E}(\vec{\zeta},\vec{\widehat{u}}(\vec{X},\vec{\zeta})) &= \int_{\Omega}\Psi(\vec{X},\vec{\zeta}, \bs{F} (\vec{\widehat{u}}(\vec{X},\vec{\zeta})) )\,\mathrm{d}\vec{X},  \quad \Psi(\vec{X}, \vec{\zeta}, \bs{F}) = \chi(\vec{X}+\vec{\zeta})\psi(\bs{F}),
\end{aligned}
\quad \forall\vec{\zeta} \in Q.
\label{problem:eq3}  
\end{equation}
By considering a fixed point~$\vec{X}$ in the energy density~$\Psi(\vec{X},\vec{\zeta},\bs{F})$ of Eq.~\eqref{problem:eq3} we see that this point is occupied by different material points for different translations~$\vec{\zeta}$. Consequently, for certain translation of the microstructure, $\vec{X}$ will be positioned inside a hole, whereas for other translations in the base material of the elastomeric matrix. Following the argumentation of~\cite{Smyshlyaev} and~\cite{Cherednichenko2004}, the effective solution is obtained as
\begin{equation}
\vec{\overline{u}}(\vec{X}) = \frac{1}{|Q|}\int_Q \vec{u}(\vec{X}, \vec{\zeta})\,\mathrm{d}\vec{\zeta},
\label{problem:eq4}  
\end{equation}
which corresponds to pointwise ensemble averaging (i.e. for each fixed~$\vec{X} \in \Omega$) over all the translated microstructures with a uniform probability density over~$Q$.

The effective solution~$\vec{\overline{u}}$ can be obtained by considering brute force calculations via DNS and discretisation of the integral in Eq.~\eqref{problem:eq4}, as reported in detail by~\cite{Ameen2018} for various scale ratios~$L/\ell$. Fig.~\ref{problem:fig2a} shows only the components~$u_2(X_2)$ of the displacement fields~$\vec{u}(\vec{X}, \vec{\zeta})$ for three vertical translations~$\vec{\zeta} \in \{\vec{0}, \frac{\ell}{2}\,\vec{e}_2, \frac{3\ell}{4}\,\vec{e}_2 \}$, and the corresponding component of the effective field~$\vec{\overline{u}}$. Due to $\vec{e}_1$-periodicity, the effective solution~$\vec{\overline{u}}$ is constant along the horizontal direction in this particular case. Recall that inside the holes, an ultra-soft linear-elastic material has been considered for interpolation purposes.

The local magnitude of the fluctuations for the ensemble~$\vec{u}(\vec{X}, \vec{\zeta})$ can be estimated pointwise as
\begin{equation}
\sigma(\vec{X}) = \frac{1}{|Q|}\int_Q \| \vec{u}(\vec{X},\vec{\zeta})-\vec{\overline{u}}(\vec{X}) \|_2 \,\mathrm{d}\vec{\zeta},
\label{eq:sigma}
\end{equation}
a quantity which will be referred to as standard deviation in what follows, shown as the red smooth curve in Fig.~\ref{problem:fig2b}. This field is again constant along the~$\vec{e}_1$ direction. The fluctuating part, approximately determined as~$[\vec{u}(\vec{X}, \vec{\zeta})-\vec{\overline{u}}(\vec{X})]/\sigma(\vec{X})$ for each translation~$\vec{\zeta}$, is plotted in Fig.~\ref{problem:fig2c} for~$\vec{\zeta} = \vec{0}$, and essentially corresponds to a microstructural buckling mode. The irregularities in the vicinity of the two horizontal edges originate from the fact that the individual fields~$\vec{u}(\vec{X}, \vec{\zeta})-\vec{\overline{u}}(\vec{X})$ as well as corresponding standard deviation~$\sigma(\vec{X})$ are close to zero there, resulting in numerical inaccuracies when dividing them. Since individual periodic as well as primitive cells of the microstructure mutually interact through the pattern, the amplitude of the mode (estimated as~$\sigma(\vec{X})$) and its spatial gradient are important for the effective behaviour of the system. Accordingly, this mode magnitude will be considered as part of the effective solution in addition to~$\vec{\overline{u}}(\vec{X})$ in the homogenization method developed in what follows.
%
%
\section{Homogenization Towards a Micromorphic Continuum}
\label{micromorphic}
Based on the insights obtained from the DNS results presented, a homogenization approach is developed in this section, which consists of three steps. First, a suitable ansatz that accurately decomposes the kinematic fields corresponding to all translated microstructures is developed in Section~\ref{decomposition}. It serves as a basis to build an approximation space over which the effective solutions are sought. In the next step, in Section~\ref{sec3_energy}, the ensemble averaged energy is introduced, the minimum of which corresponds to the average over all energies with translated microstructures considered at their solutions~$\vec{u}(\vec{X},\vec{\zeta})$, i.e.~at their minima. This average energy functional can be minimized directly with respect to the kinematic ansatz introduced in the first step, providing the closest possible approximation to the effective solution~$\vec{\overline{u}}(\vec{X})$ within the kinematic subspace considered. Using the standard arguments of the calculus of variations, a set of governing Euler--Lagrange equations along with the associated boundary conditions are finally derived and discussed in the third step in Section~\ref{sec3_ELeqn}.
%
%
\subsection{Kinematic Ansatz}
\label{decomposition}
The displacement field, common to all translated microstructures according to the definition in Eq.~\eqref{problem:eq3}, can be generally decomposed as
\begin{equation}
\vec{u}(\vec{X}, \vec{\zeta}) = \vec{v}_0(\vec{X}) + \sum_{i=1}^{n} v_i(\vec{X})\vec{\varphi}_i(\vec{X},\vec{\zeta}) + \vec{w}(\vec{X}, \vec{\zeta}).
\label{problem:eq5}
\end{equation}
Here, for fixed~$\vec{\zeta}$, $\vec{u}(\vec{X},\vec{\zeta})$ is the displacement field corresponding to a particular microstructure translated by~$\vec{\zeta}$ (i.e. the minimizer of the energy~$\mathcal{E}$ in Eq.~\eqref{problem:eq3}). For a fixed position in the specimen~$\vec{X}$, we may similarly interpret~$\vec{u}(\vec{X},\vec{\zeta})$ as the displacement at that particular point as a function of the translation vector~$\vec{\zeta}$. The vector fields~$\vec{\varphi}_i(\vec{X},\vec{\zeta})$, $i = 1, \dots, n$, are long-range spatially correlated modes characteristic of the problem morphology and deformed underlying microstructural pattern (translated by~$\vec{\zeta}$), which are in addition scaled by their magnitudes~$v_i$. Individual modes~$\vec{\varphi}_i(\vec{X},\vec{\zeta})$ are periodic and translate along with the microstructure, i.e.~$\vec{\varphi}_i(\vec{X},\vec{\zeta}) = \vec{\varphi}_i(\vec{X}+\vec{\zeta},\vec{0})$, and have zero mean. As a result of this, and of the zero mean of~$\vec{w}(\vec{X},\vec{\zeta})$ (see below), $\vec{v}_0(\vec{X})$ corresponds to the mean effective displacement field. This decomposition enables the fields~$v_i$ to `switch off' individual modes near the boundaries, for example, and thus serve as `amplitude modulators' for the individual modes~$\vec{\varphi}_i(\vec{X},\vec{\zeta})$. For a motivation of such a general decomposition see for instance~\cite{Okumura:2004}, where multiple bifurcations in foams have been studied.

For the reference problem considered in this manuscript, only one spatially correlated mode is adopted, i.e.~$n = 1$, which is defined as
\begin{equation}
\begin{aligned}
\vec{\varphi}_1(\vec{X},\vec{\zeta}) 
&= 
\frac{1}{C_1}\left[-\sin\frac{\pi}{\ell}(X_1+\zeta_1+X_2+\zeta_2) - \sin\frac{\pi}{\ell}(-X_1-\zeta_1+X_2+\zeta_2) \right]\vec{e}_1 \\
&+
\frac{1}{C_1}\left[\sin\frac{\pi}{\ell}(X_1+\zeta_1+X_2+\zeta_2) - \sin\frac{\pi}{\ell}(-X_1-\zeta_1+X_2+\zeta_2) \right]\vec{e}_2,
\end{aligned}
\label{problem:eq6}
\end{equation}
see Fig.~\ref{problem:fig2d}. In Eq.~\eqref{problem:eq6}, $C_1 = \frac{1}{|Q|}\int_Q \| \vec{\varphi}_1(\vec{X},\vec{0}) \|_2 \,\mathrm{d}\vec{X}$ is a normalizing constant ensuring that the standard deviation of~$\vec{\varphi}_1(\vec{X},\vec{\zeta})$ equals one, recall Eq.~\eqref{eq:sigma}. One then expects~$v_1$ to approximately recover the shape of the standard deviation shown in Fig.~\ref{problem:fig2b}, while~$\vec{v}_0$ is assumed to capture the trend of the mean, as depicted by the thick red line in Fig.~\ref{problem:fig2a}. The component~$v_1\vec{\varphi}_1$ essentially establishes the kinematical interactions between individual neighbouring periodic cells through the mode~$\vec{\varphi}_1$. For the present case these interactions can be interpreted as a micro rotation field, cf. Figs.~\ref{problem:fig2c} and~\ref{problem:fig2d}. The mode amplitude is therefore the highest in the bulk of the specimen, and due to the imposed constraints, equals zero at the top and bottom boundaries. The mode is moreover active only after the specimen undergoes instability, whereas in the pre-bifurcation regime it is expected to be inactive (i.e. $v_1$ vanishes). 

Finally, $\vec{w}(\vec{X}, \vec{\zeta})$ is the fast and local microfluctuation field complementary to the sum of the smooth and spatially correlated parts~$\vec{v}_0(\vec{X}) + v_1(\vec{X})\vec{\varphi}_1(\vec{X}, \vec{\zeta})$. The~$\vec{w}$ component is expected to capture fluctuations in the linear regime (in which~$v_1$ vanishes), and to compensate any inaccuracies of the mode~$\vec{\varphi}_1$ in the non-linear post-bifurcation regime. Because the split in Eq.~\eqref{problem:eq5} is not unique, additional conditions are required. Considering a fixed point in space~$\vec{X}$, both effective fields~$\vec{v}_0(\vec{X})$ and~$v_1(\vec{X})$ are constant as a function of~$\vec{\zeta}$ in Eq.~\eqref{problem:eq5}, meaning that~$\vec{w}$ should be orthogonal to a constant function in~$\vec{\zeta}$ (represented by~$\vec{v}_0$), and to~$\vec{\varphi}_1$ (which is multiplied by a constant arbitrary magnitude~$v_1$). Expressed mathematically, this means that
\begin{equation}
\begin{aligned}
\langle \vec{w}(\vec{X}, \vec{\zeta}), \vec{\varphi}_1(\vec{X}, \vec{\zeta}) \rangle_2 &= \int_Q  \vec{w}(\vec{X}, \vec{\zeta}) \cdot \vec{\varphi}_1(\vec{X}, \vec{\zeta}) \, \mathrm{d}\vec{\zeta} = 0, \\
\langle \vec{w}(\vec{X}, \vec{\zeta}), \vec{1} \rangle_2 &= \int_Q  \vec{w}(\vec{X}, \vec{\zeta}) \cdot \vec{1} \, \mathrm{d}\vec{\zeta} = 0,
\end{aligned}
\quad \forall \vec{X} \in \Omega,
\label{micromorphic:eq22}
\end{equation}
where~$\vec{1} = \vec{e}_1 + \vec{e}_2$ is a constant vector in~$\vec{\zeta}$. An extension to multiple modes~$\vec{\varphi}_i$, $i = 1, \dots, n>1$, requires orthogonality with respect to the entire vector space spanned by the considered modes.
%
%
\subsection{Energy Considerations}
\label{sec3_energy}
The energy associated with one fixed realization~$\vec{\zeta} \in Q$ of the translated microstructure reads
\begin{equation}
\mathcal{E}(\vec{\zeta},\vec{u}(\vec{X}, \vec{\zeta})) = \underset{\vec{\widehat{u}}(\vec{X}, \vec{\zeta}) \in \mathscr{U}(\Omega)}{\mbox{min}} \ \mathcal{E}(\vec{\zeta},\vec{\widehat{u}}(\vec{X}, \vec{\zeta})),
\label{micromorphic:eq1}
\end{equation}
whereas the ensemble averaged energy may be written as~\citep[see also][]{Smyshlyaev,Cherednichenko2004,Peerlings:2004}
\begin{equation}
\frac{1}{|Q|}\int_Q \mathcal{E}(\vec{\zeta},\vec{u}(\vec{X}, \vec{\zeta})) \,\mathrm{d}\vec{\zeta} = 
\frac{1}{|Q|}\int_Q \underset{\vec{\widehat{u}}(\vec{X}, \vec{\zeta}) \in \mathscr{U}(\Omega)}{\mbox{min}} \ \mathcal{E}(\vec{\zeta},\vec{\widehat{u}}(\vec{X}, \vec{\zeta})) \,\mathrm{d}\vec{\zeta} = 
\underset{\vec{\widehat{u}}(\vec{X}, \vec{\zeta}) \in \mathscr{U}(\Omega)}{\mbox{min}} \ \overline{\mathcal{E}}(\vec{\widehat{u}}(\vec{X}, \vec{\zeta})).
\label{micromorphic:eq2}
\end{equation}
Note that in Eq.~\eqref{micromorphic:eq2}, the minimization with respect to~$\vec{\widehat{u}}(\vec{X},\vec{\zeta})$ in the first equality is considered for all~$\vec{X}$ and only one particular fixed~$\vec{\zeta}$, whereas in the second equality~$\vec{\zeta}$ is considered as a continuous variable equally to~$\vec{X}$; this means effectively that~$\vec{\widehat{u}}(\vec{X},\vec{\zeta})$ is a vector function over a four-dimensional space (in the case of~$\Omega \subset \mathbb{R}^2$ and~$Q \subset \mathbb{R}^2$ as considered here). In the second equality of Eq.~\eqref{micromorphic:eq2}, the average energy functional has also been introduced as
\begin{equation}
\overline{\mathcal{E}}(\vec{\widehat{u}}(\vec{X},\vec{\zeta})) = \frac{1}{|Q|}\int_Q \int_{\Omega} \Psi(\vec{X},\vec{\zeta}, \bs{F} (\vec{\widehat{u}}(\vec{X},\vec{\zeta}))) \,\mathrm{d}\vec{X} \mathrm{d}\vec{\zeta}.
\label{micromorphic:eq3}
\end{equation}

The decomposition introduced in Eq.~\eqref{problem:eq5} is next used as a basis for the minimization of the variational homogenization problem by considering the set of admissible functions in the form
\begin{equation}
\vec{\widehat{u}}(\vec{X},\vec{\zeta}) = \vec{\widehat{v}}_0(\vec{X}) + \widehat{v}_1(\vec{X})\vec{\varphi}_1(\vec{X},\vec{\zeta}) + \vec{\widehat{w}}(\vec{X},\vec{\zeta}),
\label{micromorphic:eq21}
\end{equation}
where the present problem is limited to one mode ($n=1$) only. The problem in Eq.~\eqref{micromorphic:eq2} is then effectively replaced with
\begin{equation}
\underset{\vec{\widehat{u}}(\vec{X}, \vec{\zeta}) \in \mathscr{U}(\Omega)}{\mbox{min}} \ \overline{\mathcal{E}}(\vec{\widehat{u}}(\vec{X}, \vec{\zeta})),
\label{micromorphic:eq6}
\end{equation}
from which the kinematical functions~$\vec{v}_0(\vec{X})$, $v_1(\vec{X})$, and microfluctuation functions~$\vec{w}(\vec{X},\vec{\zeta})$ can readily be obtained as minimizers, i.e.
\begin{equation}
(\vec{v}_0(\vec{X}),v_1(\vec{X}),\vec{w}(\vec{X},\vec{\zeta})) \in \underset{\vec{\widehat{u}}(\vec{X}, \vec{\zeta}) \in \mathscr{U}(\Omega)}{\mbox{arg min}} \ \overline{\mathcal{E}}(\vec{\widehat{u}}(\vec{X}, \vec{\zeta})).
\label{micromorphic:eq7}  
\end{equation}
The minimizer~$\vec{v}_0$ results in as close an approximation to~$\vec{\overline{u}}$ as possible within the test space~$\mathscr{U}(\Omega)$ considered, i.e. within the decomposition according to Eq.~\eqref{micromorphic:eq21}. So far the exact solution for all~$\vec{\zeta}$ can be represented because~$\vec{w}$ still allows to cover the entire space~$\mathscr{U}(\Omega)$. Later, however, this space will be restricted by adopting certain constraints on~$\vec{w}$, which introduces an approximation.
%
%
\subsection{Euler--Lagrange Equations}
\label{sec3_ELeqn}
Following the standard argumentation of variational calculus, we take in the next step the G\^{a}teaux derivative of~$\overline{\mathcal{E}}$ and require it to vanish, which provides
\begin{equation}
\begin{aligned}
&
\delta\overline{\mathcal{E}}(\vec{v}_0,v_1,\vec{w};\delta\vec{v}_0,\delta v_1,\delta\vec{w}) = \left. \frac{\mathrm{d}}{\mathrm{d}h} \overline{\mathcal{E}}(\vec{v}_0+h\delta\vec{v}_0 + (v_1+h\delta v_1)\vec{\varphi}_1+\vec{w}+h\delta\vec{w})\right|_{h = 0} \\
&=
\frac{1}{|Q|}\int_Q \int_\Omega \underbrace{\frac{\partial \Psi(\vec{X},\vec{\zeta},\bs{F})}{\partial \bs{F}^\mathsf{T}}}_{\bs{P}(\vec{X},\vec{\zeta})}
:
\delta\bs{F}^\mathsf{T}(\vec{u}(\vec{X},\vec{\zeta}))\,\mathrm{d}\vec{X}\mathrm{d}\vec{\zeta} \\
&=
\frac{1}{|Q|}\int_\Omega \int_Q \bs{P}(\vec{X},\vec{\zeta}):\vec{\nabla}\delta\vec{u}(\vec{X},\vec{\zeta})\,\mathrm{d}\vec{\zeta}\mathrm{d}\vec{X} = 0,
\end{aligned}
\label{micromorphic:eq8}  
\end{equation}
where the fully-expanded definition of the displacement gradient for the considered kinematic field of Eq.~\eqref{micromorphic:eq21} reads
\begin{equation}
\vec\nabla\vec{u}(\vec{X},\vec{\zeta}) = \vec\nabla\vec{v}_0(\vec{X}) + \vec\nabla v_1(\vec{X})\vec{\varphi}_1(\vec{X},\vec{\zeta}) + v_1(\vec{X})\vec\nabla\vec{\varphi}_1(\vec{X},\vec{\zeta}) +  \vec\nabla\vec{w}(\vec{X},\vec{\zeta}),
\label{r_zeta}
\end{equation}
and, accordingly, 
\begin{equation}
\vec\nabla\delta\vec{u}(\vec{X},\vec{\zeta}) = \vec\nabla\delta\vec{v}_0(\vec{X}) + \vec\nabla \delta v_1(\vec{X})\vec{\varphi}_1(\vec{X},\vec{\zeta}) + \delta v_1(\vec{X})\vec\nabla\vec{\varphi}_1(\vec{X},\vec{\zeta}) +  \vec\nabla\delta\vec{w}(\vec{X},\vec{\zeta}).
\label{r_zeta_a}
\end{equation}

Before proceeding, let us note that one could, at this stage, make use of the divergence theorem in the last row of Eq.~\eqref{micromorphic:eq8} to obtain
\begin{equation}
\frac{1}{|Q|}\int_\Omega \int_Q (-\vec{\nabla}\cdot\bs{P}^\mathsf{T})
\cdot\delta\vec{u}\,\mathrm{d}\vec{\zeta}\mathrm{d}\vec{X} + \mbox{boundary terms} = 0,
\label{micromorphic:eq8temp}  
\end{equation}
which requires that the balance law~$\vec{\nabla}\cdot\bs{P}^\mathsf{T} = \vec{0}$ holds for each realization~$\vec{\zeta}$. This condition is satisfied only if~$\vec{w}$ is unconstrained (i.e.~the decomposition of Eq.~\eqref{micromorphic:eq21} covers the entire kinematic space~$\mathscr{U}(\Omega)$), meaning that Eqs.~\eqref{micromorphic:eq8} and~\eqref{micromorphic:eq8temp} reduce to a classical continuum. This would require computing the exact solution everywhere in~$\Omega$ for each translation~$\vec{\zeta}$. Below we will restrict the space in which~$\vec{w}$ is considered and require only that the ensemble average vanishes, meaning that only the ensemble average of the last row in Eq.~\eqref{micromorphic:eq8} is zero and that individual realizations may not be in equilibrium under constraints on~$\vec{w}$ imposed.

In order to establish ensemble averaged balance laws for the homogenized quantities, i.e. stress quantities averaged over~$Q$ and tested by variations of the effective fields~$\delta\vec{v}_0$ and~$\delta v_1$, the last row of Eq.~\eqref{micromorphic:eq8} is rewritten as
\begin{equation}
\begin{aligned}
&\frac{1}{|Q|}\int_\Omega \int_Q \bs{P}(\vec{X},\vec{\zeta}):\vec{\nabla}\delta\vec{v}_0(\vec{X})
+
\bs{P}(\vec{X},\vec{\zeta}):[\vec\nabla \delta v_1(\vec{X})\vec{\varphi}_1(\vec{X},\vec{\zeta}) + \delta v_1(\vec{X})\vec\nabla\vec{\varphi}_1(\vec{X},\vec{\zeta})] \\
&+
\bs{P}(\vec{X},\vec{\zeta}):\vec{\nabla}\vec{w}(\vec{X},\vec{\zeta}) \,\mathrm{d}\vec{\zeta} \mathrm{d}\vec{X} \\
&=
\int_\Omega \Bigg\{\underbrace{\frac{1}{|Q|}\int_Q \bs{P}(\vec{X},\vec{\zeta})\,\mathrm{d}\vec{\zeta}}_{\bs{P}_1(\vec{X})}:\vec{\nabla}\delta\vec{v}_0(\vec{X})
+
\underbrace{\frac{1}{|Q|}\int_\Omega \bs{P}^\mathsf{T}(\vec{X},\vec{\zeta})\cdot\vec{\varphi}_1(\vec{X},\vec{\zeta})\,\mathrm{d}\vec{\zeta}}_{\vec{P}_3(\vec{X})}\cdot\vec{\nabla}\delta v_1(\vec{X}) \\
&+
\underbrace{\frac{1}{|Q|}\int_\Omega \bs{P}(\vec{X},\vec{\zeta}):\vec{\nabla}\vec{\varphi}_1(\vec{X},\vec{\zeta})\,\mathrm{d}\vec{\zeta}}_{P_2(\vec{X})}\,\delta v_1(\vec{X})
+
\frac{1}{|Q|}\int_Q\bs{P}(\vec{X},\vec{\zeta}):\vec{\nabla}\vec{w}(\vec{X},\vec{\zeta})\,\mathrm{d}\vec{\zeta}\Bigg\} \, \mathrm{d}\vec{X} = 0.
\end{aligned}
\label{micromorphic:eq8b}  
\end{equation}
In compact form, Eq.~\eqref{micromorphic:eq8b} reads
\begin{equation}
\begin{aligned}
&\int_\Omega \Bigg\{ \bs{P}_1(\vec{X}):\vec{\nabla}\delta\vec{v}_0(\vec{X}) + \vec{P}_3(\vec{X})\cdot\vec{\nabla}\delta v_1(\vec{X}) + P_2(\vec{X})\delta v_1(\vec{X}) \\
&+
\frac{1}{|Q|}\int_Q\bs{P}(\vec{X},\vec{\zeta}):\vec{\nabla}\vec{w}(\vec{X},\vec{\zeta}) \,\mathrm{d}\vec{\zeta} \Bigg\}\mathrm{d}\vec{X} = 0,
\end{aligned}
\label{micromorphic:eq8d}  
\end{equation}
which, with the help of the divergence theorem applied to~$\vec{\nabla}\delta\vec{v}_0$, $\vec{\nabla}\delta v_1$, and~$\vec{\nabla}\delta\vec{w}$, can be rewritten as
\begin{equation}
\begin{aligned}
&\int_\Omega \Bigg\{[-\vec{\nabla}\cdot\bs{P}_1^\mathsf{T}(\vec{X})]\cdot\delta\vec{v}_0(\vec{X}) + [P_2(\vec{X})-\vec{\nabla}\cdot\vec{P}_3(\vec{X})]\delta v_1(\vec{X}) \\
&+
\frac{1}{|Q|}\int_Q [-\vec{\nabla}\cdot\bs{P}^\mathsf{T}(\vec{X},\vec{\zeta})]\cdot\delta\vec{w}(\vec{X},\vec{\zeta})
\,\mathrm{d}\vec{\zeta} \Bigg\} \mathrm{d}\vec{X} 
+
\int_{\Gamma_\mathrm{N}} \delta\vec{v}_0(\vec{X})\cdot\bs{P}_1(\vec{X})\cdot\vec{N}(\vec{X})
\,\mathrm{d}\vec{X} \\
&+
\int_{\Gamma_\mathrm{N}} \delta v_1(\vec{X})\vec{P}_3(\vec{X})\cdot\vec{N}(\vec{X})
\,\mathrm{d}\vec{X}
+
\frac{1}{|Q|}\int_{\Gamma_\mathrm{N}} \int_Q \delta\vec{w}(\vec{X},\vec{\zeta})\cdot\bs{P}(\vec{X},\vec{\zeta})\cdot\vec{N}(\vec{X})
\,\mathrm{d}\vec{\zeta} \mathrm{d}\vec{X} = 0,
\end{aligned}
\label{micromorphic:eq8e}  
\end{equation}
where~$\Gamma_\mathrm{N}$ denotes the free part of the domain boundary~$\partial\Omega$, and~$\vec{N}$ the unit outer normal to~$\partial\Omega$ in the reference configuration. Recall, however, that prescribed tractions have been omitted in the definition of the ensemble averaged energy of Eq.~\eqref{micromorphic:eq3}, and that only essential boundary conditions on the~$\Gamma_\mathrm{D}$ part of the boundary along with periodic boundary conditions have been considered.

Making use of the localization argument in space as well as in translations (because~$\vec{w}$ is a space-translation quantity), Eq.~\eqref{micromorphic:eq8e} gives the following set of governing Euler--Lagrange equations
\begin{equation}
\delta\vec{v}_0:\ \left\{
\begin{aligned}
\vec{\nabla}\cdot\bs{P}_1^\mathsf{T} &= \vec{0}, \quad \mbox{in}\ \Omega,\\
\bs{P}_1\cdot\vec{N} &= \vec{0}, \quad \mbox{on}\ \Gamma_\mathrm{N}, \\
\end{aligned}\right.
\quad\delta v_1:\ \left\{
\begin{aligned}
\vec{\nabla}\cdot \vec{P}_3-P_2 &= 0, \quad \mbox{in}\ \Omega,\\
\vec{P}_3\cdot\vec{N} &= 0, \quad \mbox{on}\ \Gamma_\mathrm{N}, \\
\end{aligned}\right.
\label{micromorphic:eq9}
\end{equation}
\begin{equation}
\quad\delta\vec{w}:\ \left\{
\begin{aligned}
\vec{\nabla}\cdot\bs{P}^\mathsf{T} &= \vec{0}, \quad \mbox{in}\ \Omega, \\
\bs{P}\cdot\vec{N} &= \vec{0}, \quad \mbox{on}\ \Gamma_\mathrm{N}, 
\end{aligned}\right. \quad \forall \vec{\zeta} \in Q,
\label{micromorphic:eq10}
\end{equation}
where the averaged stress quantities~$\bs{P}_1(\vec{X})$, $P_2(\vec{X})$, and~$\vec{P}_3(\vec{X})$, have been introduced as
\begin{equation}
\begin{aligned}
\bs{P}_1(\vec{X}) &= \frac{1}{|Q|}\int_Q \bs{P}(\vec{X},\vec{\zeta}) \, \mathrm{d}\vec{\zeta}, \quad
P_2(\vec{X}) = \frac{1}{|Q|}\int_Q \bs{P}(\vec{X},\vec{\zeta}):\vec{\nabla}\vec{\varphi}_1(\vec{X},\vec{\zeta}) \, \mathrm{d}\vec{\zeta}, \\
\vec{P}_3(\vec{X}) &= \frac{1}{|Q|}\int_Q \bs{P}^\mathsf{T}(\vec{X},\vec{\zeta})\cdot\vec{\varphi}_1(\vec{X},\vec{\zeta}) \, \mathrm{d}\vec{\zeta}.
\end{aligned}
\label{micromorphic:eq13}  
\end{equation}
Although the stress quantity~$\bs{P}_1$ is a simple ensemble average and its balance law in Eq.~(\ref{micromorphic:eq9}a) has the standard form, the balance law is not standard as~$\bs{P}_1$ depends also on~$v_1$ and~$\vec{\nabla}\vec{v}_1$. Through these two terms the balance law in Eq.~(\ref{micromorphic:eq9}a) couples to the non-standard equilibrium Eq.~(\ref{micromorphic:eq9}b), expressed in terms of the two non-standard stress quantities~$P_2$ and~$\vec{P}_3$. Since all homogenized stresses depend on (gradients) of~$\vec{v}_0$ and~$v_1$, a micromorphic continuum emerges. So far, all stresses depend on~$\vec{\nabla}\vec{w}$ as well.

As Eq.~\eqref{micromorphic:eq9} has been derived for the ensemble averaged quantities~$\vec{v}_0$ and~$v_1$, it depends on the coordinate vector~$\vec{X}$ only. In contrast to that, $\vec{w}$ and its variation~$\delta\vec{w}$ are also functions of~$\vec{\zeta}$, and hence~$\delta\vec{w}$ cannot be taken outside the integral over~$Q$ in the derivation of Eq.~\eqref{micromorphic:eq8e}. This effectively means that~$\vec{w}$ needs to be solved in Eq.~\eqref{micromorphic:eq10} for each translation~$\vec{\zeta}$ separately.

Unlike the natural boundary conditions expressed in terms of~$\bs{P}_1$, $\vec{P}_2$, and obtained from the minimization procedure of the ensemble averaged energy (cf. Eqs.~\eqref{micromorphic:eq8e} and~\eqref{micromorphic:eq9}), the essential boundary conditions for all coarse fields~$\vec{v}_0$ and, in general, $v_i$ can be derived directly from the considerations of individual translations, recall Section~\ref{full_problem} and Fig.~\ref{problem:fig2}. This may generally not be as straightforward for other homogenization techniques based on, e.g., volume averaging. In the case of the developed micromorphic homogenization technique it is clear, however, that when the displacement field is prescribed on a part of the domain boundary~$\Gamma_\mathrm{D}\subset\partial\Omega$, i.e.~$\vec{u}(\vec{X},\vec{\zeta}) = \vec{u}_\mathrm{D}(\vec{X})$ on~$\Gamma_\mathrm{D}$, all fluctuations~$v_i(\vec{X})\vec{\varphi}_i(\vec{X},\vec{\zeta})$ need to vanish there for all translations~$\vec{\zeta}\in Q$. This effectively means that all coarse modulating functions vanish at this part of the boundary, i.e.~$v_i(\vec{X}) = 0$ on~$\Gamma_\mathrm{D}$, whereas~$\vec{v}_0(\vec{X}) = \vec{u}_\mathrm{D}(\vec{X})$ on~$\Gamma_\mathrm{D}$ captures any prescribed displacements.

In total three unknown fields are considered in Eqs.~\eqref{micromorphic:eq9} and~\eqref{micromorphic:eq10}, namely~$\vec{v}_0(\vec{X})$, $v_1(\vec{X})$, and~$\vec{w}(\vec{X},\vec{\zeta})$, which will be discretised using the finite element method. Assuming that~$\vec{v}_0(\vec{X})$ and~$v_1(\vec{X})$ are slowly varying effective fields, their spatial approximation or discretisation may be relatively coarse, considered at the level of the entire specimen~$L$. On the contrary, $\vec{w}(\vec{X},\vec{\zeta})$ is a rapidly oscillating field, which needs to be resolved finely at the microstructural level~$\ell$ for each translation~$\vec{\zeta} \in Q$. In order not to solve for~$\vec{w}(\vec{X},\vec{\zeta})$ globally inside the entire domain~$\Omega$ (which would effectively correspond to the DNS solutions for all translated microstructures), two approaches are presented below:
\begin{itemize}
	\item First, $\vec{w}(\vec{X},\vec{\zeta})$ is neglected and considered as a truncation error in the closed-form homogenization approach detailed in Section~\ref{simplified}. This is only a reasonable assumption if a high approximation quality can be achieved with the decomposition~$\vec{v}_0+v_1\vec{\varphi}_1$, i.e. when~$\vec{\varphi}_1$ captures most of the significant fluctuations present.
	\item Second, $\vec{w}(\vec{X},\vec{\zeta})$ is computed only locally inside a periodic cell associated with each coarse-scale Gauss integration point for only one translation, reducing to a computational homogenization approach. This is discussed in more detail in Section~\ref{fe2}.
\end{itemize}
%
%
\section{Closed-Form Homogenization Approach}
\label{simplified}
The first approach adopted neglects the microfluctuation field~$\vec{w}$ (i.e.~$\vec{w}$ is constrained to~$\vec{w}(\vec{X},\vec{\zeta}) \equiv \vec{0}$), assuming that the leading terms~$\vec{v}_0+v_1\vec{\varphi}$ in Eq.~\eqref{micromorphic:eq21} approximate the macro- as well as micro-kinematics sufficiently accurately. Upon such a simplification, the governing Eqs.~\eqref{micromorphic:eq9} and~\eqref{micromorphic:eq10} effectively reduce to Eq.~\eqref{micromorphic:eq9} for the homogenized quantities only, along with the definitions of the corresponding conjugate stresses in Eq.~\eqref{micromorphic:eq13}. Consequently, the resulting problem can readily be discretised and solved as follows.
%
%
\subsection{Discretisation of the Governing Equations}
\label{simplified:discretization}
A macroscopic discretisation is introduced, and both effective fields are expressed in terms of shape functions as
\begin{equation}
\vec{v}_0(\vec{X}) \approx \bs{\mathsf{N}}_0(\vec{X})\underline{v}_0, \quad v_1(\vec{X}) \approx \bs{\mathsf{N}}_1(\vec{X})\underline{v}_1,
\label{simplified:discretization:eq1}
\end{equation}
where~$\underline{v}_0$ and~$\underline{v}_1$ are columns storing nodal values of the respective fields, and~$\bs{\mathsf{N}}_0(\vec{X})$, $\bs{\mathsf{N}}_1(\vec{X})$, are matrices storing the individual shape functions. Spatial derivatives are then expressed as
\begin{equation}
\vec{\nabla}\vec{v}_0(\vec{X}) \approx \bs{\mathsf{B}}_0(\vec{X})\underline{v}_0, \quad \vec{\nabla}v_1(\vec{X}) \approx \bs{\mathsf{B}}_1(\vec{X})\underline{v}_1,
\label{simplified:discretization:eq2}
\end{equation}
where~$\bs{\mathsf{B}}_0(\vec{X})$ and~$\bs{\mathsf{B}}_1(\vec{X})$ are standard matrices of the shape functions' spatial derivatives. The variations~$\delta\vec{v}_0$, $\delta v_1$, and their spatial derivatives are approximated in the same way.

Making use of the first variation of the average energy functional~$\overline{\mathcal{E}}$ in Eq.~\eqref{micromorphic:eq8d} and the definitions of homogenized stresses in Eq.~\eqref{micromorphic:eq13}, one can write for the column matrix of internal forces
\begin{equation}
\underline{0} = \underline{f} = \left[\begin{array}{c}\underline{f}_0\\\underline{f}_1\end{array}\right],
\quad \mbox{where}\quad\left\{ \quad
\begin{aligned}
\underline{f}_0 &= \Aop_{e=1}^{n_\mathrm{e}} \int_{\Omega_e} \bs{\mathsf{B}}_0^\mathsf{T}(\vec{X})\underline{P}_1(\vec{X})\,\mathrm{d}\vec{X}, \\
\underline{f}_1 &= \Aop_{e=1}^{n_\mathrm{e}} \int_{\Omega_e} \bs{\mathsf{B}}_1^\mathsf{T}(\vec{X})\underline{P}_3(\vec{X}) + \bs{\mathsf{N}}_1^\mathsf{T}(\vec{X})P_2(\vec{X})\,\mathrm{d}\vec{X},
\end{aligned}\right.
\label{simplified:discretization:eq3}
\end{equation}
where~$\mathsf{A}$ denotes the assembly operator over~$n_\mathrm{e}$ elements with spatial subdomains~$\Omega_e$. For the average energy itself, one can write
\begin{equation}
\overline{\mathcal{E}} \approx \sum_{e=1}^{n_\mathrm{e}} \int_{\Omega_e} \overline{\Psi}(\vec{X})\,\mathrm{d}\vec{X} \approx  \sum_{i_\mathrm{g}=1}^{n_\mathrm{g}}w_{i_\mathrm{g}}J_{i_\mathrm{g}}\overline{\Psi}_{i_\mathrm{g}}.
\label{simplified:discretization:eq4}
\end{equation}
In both Eqs.~\eqref{simplified:discretization:eq3} and~\eqref{simplified:discretization:eq4}, the integrals over individual elements~$\Omega_e$ are carried out numerically in the standard way using a Gauss integration rule, as indicated in Eq.~\eqref{simplified:discretization:eq4}. Here, the total average energy is computed as a sum over all integration points, where~$w_{i_\mathrm{g}}$ is the integration weight, $J_{i_\mathrm{g}}$ is associated Jacobian, and~$\overline{\Psi}_{i_\mathrm{g}}$ denotes the average energy density associated with the considered integration point~$i_\mathrm{g}$.

Note that when the microfluctuation field~$\vec{w}$ is not present, all average conjugate quantities~$\bs{P}_1$, $P_2$, $\vec{P}_3$, and the average energy density~$\overline{\Psi}$, can be computed by simply sampling the corresponding constitutive law (no solution at the `microscale' is required). Let us consider a fixed Gauss integration point~$i_\mathrm{g}$, in which the coarse fields~$\vec{\nabla}\vec{v}_0$, $v_1$, and~$\vec{\nabla}v_1$ are known. Because also a fixed position is assumed, no further function dependencies on~$\vec{X}$ need to be considered. As a consequence, the deformation gradient becomes an explicit function of~$\vec{\zeta}$ only, i.e.
\begin{equation}
\bs{F}^\mathsf{T}(\vec{\zeta}) = \bs{I} + \vec{\nabla}\vec{v}_0 + \vec{\nabla}v_1\vec{\varphi}_1(\vec{\zeta}) + v_1\vec{\nabla}\vec{\varphi}_1(\vec{\zeta}),
\label{simplified:discretization:eq5}
\end{equation}
where~$\vec{\varphi}_1(\vec{\zeta})$ is an a priori known vector field, recall Eq.~\eqref{problem:eq6}. The deformation gradient Eq.~\eqref{simplified:discretization:eq5} can thus directly be substituted into the adopted constitutive law and integrated over~$Q$, i.e.
\begin{equation}
\overline{\Psi} = \frac{1}{|Q|}\int_Q \Psi(\vec{0},\vec{\zeta},\bs{F}(\vec{\zeta})) \, \mathrm{d}\vec{\zeta}, \quad \bs{P}_1 = \frac{1}{|Q|}\int_Q \bs{P}(\vec{0},\vec{\zeta},\bs{F}(\vec{\zeta})) \, \mathrm{d}\vec{\zeta}.
\label{simplified:discretization:eq6}
\end{equation}
Similar expressions also hold for the remaining homogenized quantities~$P_2$ and~$\vec{P}_3$. In Eq.~\eqref{simplified:discretization:eq6} we have, without any loss of generality, considered the position of the Gauss integration point as~$\vec{X} = \vec{0}$.

For the numerical results presented below in Section~\ref{simplified:results}, piece-wise linear shape functions have been used with a one-point integration scheme, for both effective fields~$\vec{v}_0$ and~$v_1$. Because the example considered is periodic in~$\vec{e}_1$ direction, the effective problem reduces to be one-dimensional only, in which all averaged quantities are constant along the~$\vec{e}_1$ direction. Although a standard numerical Newton solver can be used to minimize~$\overline{\mathcal{E}}$, or in other words to equilibrate~$\underline{f}$, for convenience and for brevity a quasi-Newton solver has been used instead. This minimization technique does not require any evaluations of the Hessian of the average energy~$\overline{\mathcal{E}}$ (or in other words the tangent stiffness matrix associated with~$\underline{f}$), but more iterations are usually required.
%
%
\subsection{Results}
\label{simplified:results}
\begin{figure}
	\centering
	\includegraphics[scale=1]{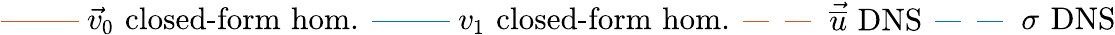}\vspace{-0.5em}\\
    \subfloat[\stackunder{$L/\ell = 5$,}{$u_\mathrm{D}/L = 0.02$}]{\includegraphics[scale=1]{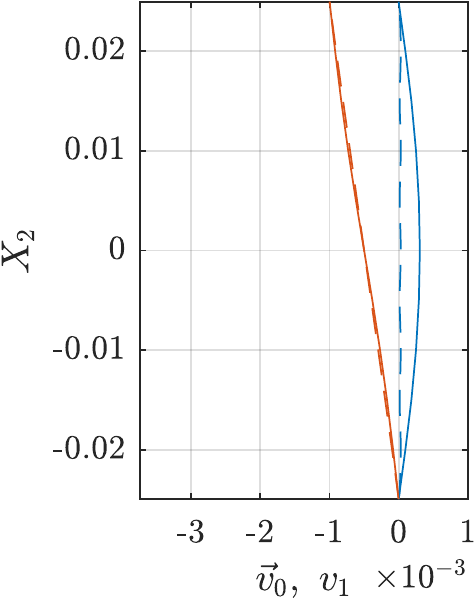}\label{simplified:results:fig1c}}	
    \subfloat[\stackunder{$L/\ell = 5$,}{$u_\mathrm{D}/L = 0.075$}]{\includegraphics[scale=1]{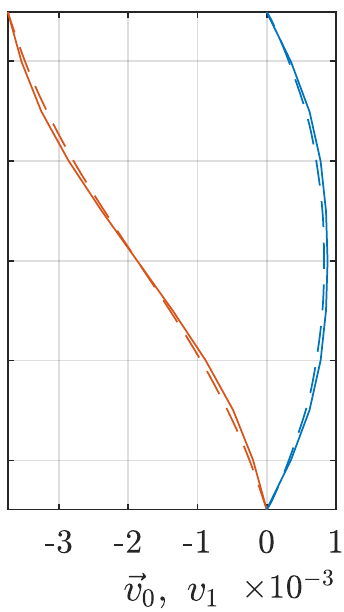}\label{simplified:results:fig1a}}	
	\subfloat[\stackunder{$L/\ell = 12$,}{$u_\mathrm{D}/L = 0.02$}]{\includegraphics[scale=1]{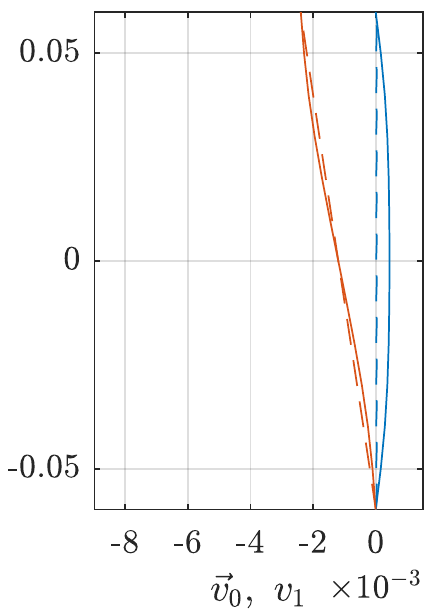}\label{simplified:results:fig1d}}
	\subfloat[\stackunder{$L/\ell = 12$,}{$u_\mathrm{D}/L = 0.075$}]{\includegraphics[scale=1]{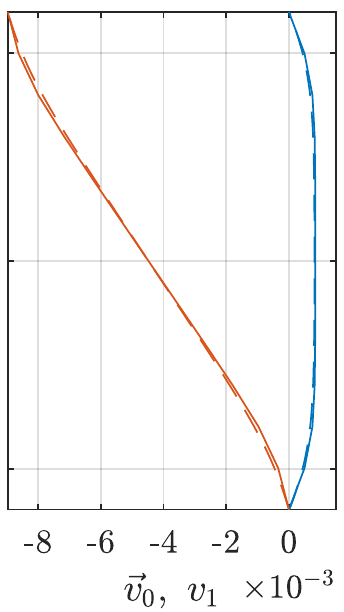}\label{simplified:results:fig1b}}	
	\caption{The effective fields~$\vec{v}_0$ (vertical component) and~$v_1$ obtained from the closed-form homogenization approach in comparison with the DNS results for scale ratio~$L/\ell = 5$ in~(a) and~(b), and~$12$ in~(c) and~(d). Two levels of nominal strain, corresponding to the linear ($u_\mathrm{D}/L = 0.02$) and post-bifurcation ($u_\mathrm{D}/L = 0.075$) regimes, are shown.}
	\label{simplified:results:fig1}
\end{figure}

The effective solutions~$\vec{v}_0(X_2)$ and~$v_1(X_2)$ obtained for two scale ratios~$L/\ell = 5$ and~$L/\ell = 12$ and for applied nominal strains of~$u_\mathrm{D}/L = 0.02$ and~$0.075$ are compared against the corresponding characteristics of the respective DNS solutions in Fig.~\ref{simplified:results:fig1}. Here we can see that the achieved accuracy in terms of kinematics is encouraging, especially in the post-bifurcation regime (Figs.~\ref{simplified:results:fig1a} and~\ref{simplified:results:fig1b}). The boundary layers are captured accurately, scaling well with the adopted scale ratio~$L/\ell$. The plateau of~$v_1$ in the bulk region, especially for the scale ratio~$12$, is also well reflected. The small differences between~$v_1$ and the standard deviation~$\sigma$ are explained by the fact that the two quantities are not directly comparable (recall that whereas~$\sigma$ includes the effect of~$\vec{w}$, $v_1$ does not). For both homogenized solutions in the pre-bifurcation regime, the term~$v_1\vec{\varphi}_1$ tries to represent the effect of~$\vec{w}$, mimicking a buckled state in which~$v_1 \neq 0$ (Figs.~\ref{simplified:results:fig1c} and~\ref{simplified:results:fig1d}); this is a result of the overly stiff (over-constrained) representation, as explained in the next paragraph.

\begin{figure}
	\centering
	\includegraphics[scale=1]{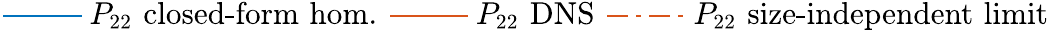}\vspace{-0.5em}\\
    \subfloat[$u_\mathrm{D}/L = 0.02$]{\includegraphics[scale=1]{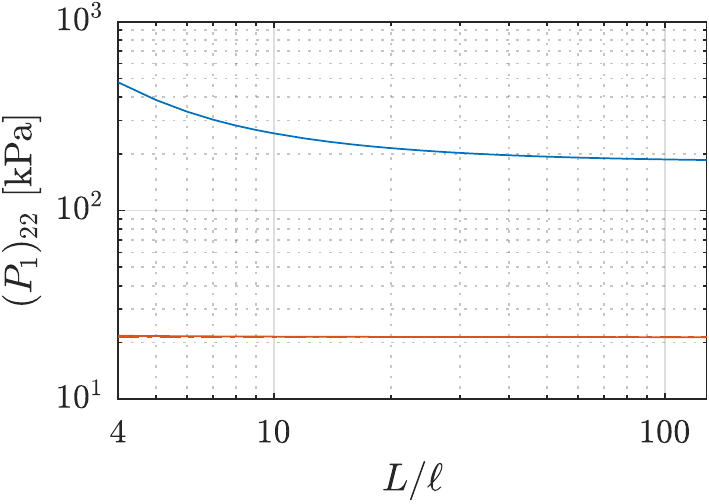}\label{simplified:results:fig3a}}\hspace{1em}
	\subfloat[$u_\mathrm{D}/L = 0.075$]{\includegraphics[scale=1]{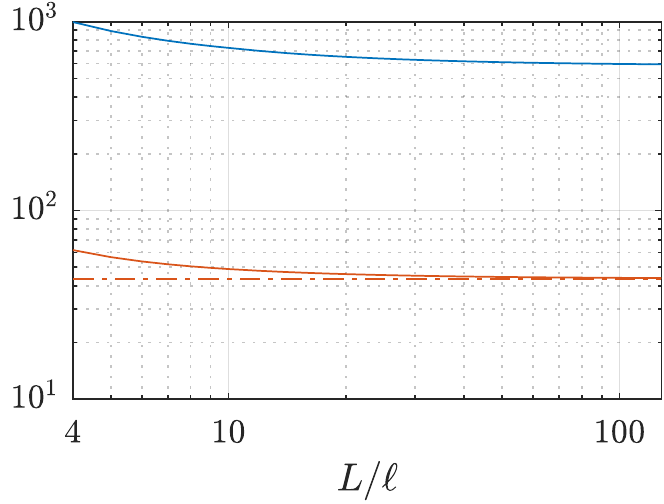}\label{simplified:results:fig3b}}
	\caption{Nominal stress~$(P_1)_{22}$ as a function of scale ratio~$L/\ell$ as obtained from the closed-form homogenization and the DNS for two applied nominal strain levels of~(a) $u_\mathrm{D}/L = 0.02$ (in the linear regime) and~(b) $u_\mathrm{D}/L = 0.075$ (post-bifurcation regime).}
	\label{simplified:results:fig3}
\end{figure}

Although the kinematic fields are relatively acceptable, especially in the post-bifurcation regime, the corresponding conjugate quantities are severely inaccurate, see Fig.~\ref{simplified:results:fig3} where a logarithmic scale has been adopted on the vertical axis to allow for comparison. Here, the (mean) vertical stress component~$(P_1)_{22}$ which results from the applied deformation as a function of the scale ratio~$L/\ell$ is shown. In the homogenized solution this quantity is simply the outcome of the numerical analysis, whereas in the DNS case it has been obtained by brute force averaging over all realizations. The dependence of the results on the scale ratio~$L/\ell$ originates from restricting the pattern transformation in a boundary layer (thickness of which scales with~$\ell$) close to the top and bottom boundaries. In the limit~$L/\ell \rightarrow \infty$ this influence becomes negligible and a scale-independent limit is reached. The closed-form homogenized solution is able to qualitatively capture this trend in the post-bifurcation regime, although the values differ by a factor of~$15$ (Fig.~\ref{simplified:results:fig3b}). Pre-buckling, however, the size effect exhibited by the closed-form homogenized solution is artificial (and wrong), as almost no dependence on the scale ratio is observed for the DNS results, see Fig.~\ref{simplified:results:fig3a}. This is due to the fact that actually almost all homogenized solutions buckle already even for applied strains as small as~$u_\mathrm{D}/L = 0.01$. As mentioned above, premature bifurcation occurs to relax the energy through the long-range correlated field~$v_1\vec{\varphi}_1$ to compensate for the neglected microfluctuation field~$\vec{w}$. This is the result of overconstraining~$\vec{w}$, which leads to an overly stiff system compared to the DNS. Note that one option to improve on this is to consider more modes~$\vec{\varphi}_i$ with shorter wavelengths, capable of approximating the vector field~$\vec{w}$ in a better way. This approach is not considered in what follows, and a computational scheme is adopted instead.
%
%
\section{Computational Homogenization Approach}
\label{fe2}
In spite of the reasonable effective post-bifurcation kinematic fields obtained under the constraint~$\vec{w} \equiv \vec{0}$, Section~\ref{simplified:results} convincingly showed that the microfluctuation field~$\vec{w}$ needs to be accounted for. As briefly mentioned at the end of Section~\ref{micromorphic}, the difficulty with~$\vec{w}$ is twofold: (i)~it is defined on the entire (full-scale) problem domain~$\Omega$, meaning that one has to basically solve the full-scale problem to obtain it, and~(ii) one needs to do this for every realization (translation of the microstructure). Hence, in order to avoid resorting to DNS type of simulations, additional approximations on~$\vec{w}$ need to be made to simplify the multi-scale ansatz.

A first approximation which we propose here is to compute (for each translation) the microfluctuation field~$\vec{w}$ on a~$2\ell \times 2\ell$ periodic cell around the point of interest (e.g.~a macroscopic integration point) instead of on the full scale problem domain~$\Omega$. This means that boundary conditions have to be applied on the boundary of that periodic cell instead of on the outer boundary of the problem domain. (Note that near the specimen boundary~$\partial\Omega$, part of this periodic cell may be outside of the problem domain~$\Omega$.) To achieve this, two assumptions are made: 
\begin{itemize}
	\item On the periodic cell, the effective fields~$\vec{v}_0$ and~$v_1$ vary only slowly and hence they can be approximated by a linear expansion around the point of interest (i.e. the centre of the periodic cell);
	\item The remainder~$\vec{w}$ is periodic over the periodic cell.
\end{itemize}
These conditions, together, allow us to estimate the energy density at a point~$\vec{X}$ for every translation~$\vec{\zeta}$.

A second approximation is made in order to avoid the computation of~$\vec{w}$ for every translation~$\vec{\zeta}$. Hereto, we estimate the energy density in a point~$\vec{X}$ for all the translations from the energy density distribution in the neighbourhood of that point in one realization only.

All considered assumptions are elaborated formally in Section~\ref{assumptions}. Approximations of the average energy along with its minimization leading to the governing equations are further discussed in Section~\ref{approx_energy}, whereas the overall computational framework is summarized in Section~\ref{computational_framewrok}. For results obtained with this framework we refer to Section~\ref{results}.
%
%
\subsection{Assumptions and Approximations}
\label{assumptions}
Using a Taylor series expansion and assuming that the~$\vec{v}_0$ and~$v_1$ fields vary spatially at a much coarser scale than~$\vec{\varphi}_1$ and~$\vec{w}$, the first assumption implies that the effective fields can be approximated in a small neighbourhood of a point~$\vec{X}$ (spanned by~$\Delta\vec{X}$) as
\begin{equation}
\begin{aligned}
\vec{v}_0(\vec{X} + \Delta\vec{X}) &= \vec{v}_0(\vec{X}) + \Delta\vec{X}\cdot\vec{\nabla}\vec{v}_0(\vec{X}) + \mathcal{O}(\|\Delta\vec{X}\|^2), \\
v_1(\vec{X} + \Delta\vec{X}) &= v_1(\vec{X}) + \Delta\vec{X}\cdot\vec{\nabla}v_1(\vec{X}) + \mathcal{O}(\|\Delta\vec{X}\|^2).
\end{aligned}
\label{fe2:eq1}
\end{equation}
The microstructural components of the ansatz Eq.~\eqref{problem:eq5}, on the other hand, fluctuate at a fine scale compared to~$\vec{v}_0$ and~$v_1$ and hence, they need to be evaluated exactly as
\begin{equation}
\vec{\varphi}_1(\vec{X}+\Delta\vec{X},\vec{\zeta}), \quad\mbox{and}\quad \vec{w}(\vec{X}+\Delta\vec{X},\vec{\zeta}).
\label{fe2:eq3}
\end{equation}
Substituting Eqs.~\eqref{fe2:eq1} and~\eqref{fe2:eq3} into Eq.~\eqref{micromorphic:eq21} provides an expression for the approximate displacement field in the form
\begin{equation}
\begin{aligned}
\vec{u}(\vec{X}+\Delta\vec{X},\vec{\zeta}) \approx \vec{\widetilde{u}}(\vec{X},\Delta\vec{X},\vec{\zeta}) &= \vec{v}_0(\vec{X}) + \Delta\vec{X}\cdot\vec{\nabla}\vec{v}_0(\vec{X}) \\
&+
[v_1(\vec{X}) + \Delta\vec{X}\cdot\vec{\nabla}v_1(\vec{X})]\vec{\varphi}_1(\vec{X}+\Delta\vec{X},\vec{\zeta}) \\
&+
\vec{w}(\vec{X}+\Delta\vec{X},\vec{\zeta}),
\end{aligned}
\label{fe2:eq2}
\end{equation}
with corresponding deformation gradient
\begin{equation}
\begin{aligned}
\bs{F}^\mathsf{T}(\vec{\widetilde{u}}(\vec{X},\Delta\vec{X},\vec{\zeta})) - \bs{I} &= \vec{\nabla}_\mathrm{m}\vec{\widetilde{u}}(\vec{X},\Delta\vec{X},\vec{\zeta}) = \vec{\nabla}\vec{v}_0(\vec{X}) + \vec{\nabla}v_1(\vec{X})\vec{\varphi}_1(\vec{X}+\Delta\vec{X},\vec{\zeta}) \\
&+
[v_1(\vec{X}) + \Delta\vec{X}\cdot\vec{\nabla}v_1(\vec{X})]\vec{\nabla}_\mathrm{m}\vec{\varphi}_1(\vec{X}+\Delta\vec{X},\vec{\zeta}) + \vec{\nabla}_\mathrm{m}\vec{w}(\vec{X}+\Delta\vec{X},\vec{\zeta}).
\end{aligned}
\label{fe2:eq4c}
\end{equation}
The differentiation with respect to a (fine scale) variable~$\Delta\vec{X}$ in Eq.~\eqref{fe2:eq4c}, i.e.~$\vec{\nabla}_\mathrm{m} = \partial/\partial\Delta X_i$, is valid because within the small neighbourhood considered, the global position~$\vec{X}$ is kept fixed. 
\begin{figure}
	\centering
	\def\svgwidth{0.7\textwidth}
	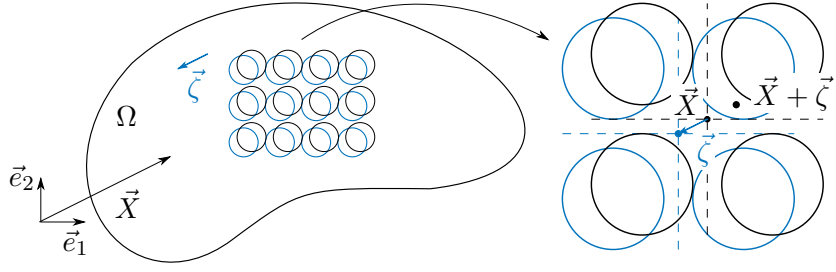
	\caption{A sketch showing a point~$\vec{X}$ in the reference microstructure~$ \vec{\zeta} = \vec{0} $ (in black) and a microstructure translated by~$\vec{\zeta}$ (in blue). The point~$\vec{X}$ in the translated microstructure is microstructurally (or pattern-wise) at the same position as the point~$\vec{X}+\vec{\zeta}$ in the reference microstructure, although these two points experience slightly different effective fields~$\vec{v}_0$ and~$v_1$.}
	\label{derivation:fig5}
\end{figure}

In the second step, two realizations as sketched in Fig.~\ref{derivation:fig5} are considered. The \textit{reference microstructure} is defined as~$\vec{\zeta} = \vec{0}$ (in black), whereas the \emph{translated microstructure} is defined as~$\vec{\zeta} \neq \vec{0}$ (in blue). A point~$\vec{X}$ in the microstructure translated by~$\vec{\zeta}$ has exactly the same relative microstructural position as the point~$\vec{X}+\vec{\zeta}$ in the reference microstructure, as indicated in the right part of Fig.~\ref{derivation:fig5}. This means that these two points are microstructurally (or pattern-wise) identical, although they are experiencing slightly different underlying effective fields~$\vec{v}_0$ and~$v_1$. Using this second assumption, the constitutive law of a point~$\vec{X}$ in a translated realization exactly corresponds to that of a point~$\vec{X}+\vec{\zeta}$ in the reference realization, i.e.
\begin{equation}
\Psi(\vec{X},\vec{\zeta},\widehat{\bs{F}}) = \Psi(\vec{X} + \vec{\zeta},\vec{0},\widehat{\bs{F}}),
\label{fe2:eq4a}
\end{equation}
where an arbitrary deformation gradient~$\widehat{\bs{F}}$ has been considered. Next, a similar procedure is applied to the deformation gradient at a point~$\vec{X}$ for the translated microstructure, i.e.~$\bs{F}(\vec{u}(\vec{X},\vec{\zeta}))$, which is approximated by the deformation gradient in the corresponding position~$\vec{X}+\vec{\zeta}$ of the reference microstructure. To this end, use of the Taylor-expanded deformation gradient in Eq.~\eqref{fe2:eq4c} is made, which provides
\begin{equation}
\bs{F}(\vec{u}(\vec{X},\vec{\zeta})) \approx \bs{F}(\vec{\widetilde{u}}(\vec{X},\vec{0},\vec{\zeta})) \approx \bs{F}(\vec{\widetilde{u}}(\vec{X},\vec{\zeta},\vec{0})).
\label{fe2:eq4b}
\end{equation}
In the first approximation of Eq.~\eqref{fe2:eq4b}, the exact deformation gradient at a position~$\vec{X}$ and microstructure translated by~$\vec{\zeta}$ has been replaced by the Taylor-expanded one from Eq.~\eqref{fe2:eq4c} (as the point~$\vec{X}$ is considered, $\Delta\vec{X} = \vec{0}$ while~$\vec{\zeta} \neq \vec{0}$), whereas in the second approximation of Eq.~\eqref{fe2:eq4b} the corresponding point~$\vec{X}+\vec{\zeta}$ in the reference microstructure is used (i.e.~the translation vector~$\vec{\zeta}$ effectively becomes a position vector relative to~$\vec{X}$, replacing~$\Delta\vec{X}$ in Eq.~\eqref{fe2:eq4c}). Translating the constitutive law (Eq.~\eqref{fe2:eq4a}) while employing the kinematic approximation of the deformation gradient (Eqs.~\eqref{fe2:eq4c} and~\eqref{fe2:eq4b}), one can finally estimate the energy density at a point~$\vec{X}$ in the microstructure translated by~$\vec{\zeta}$ as
\begin{equation}
\Psi(\vec{X},\vec{\zeta},\bs{F}(\vec{u}(\vec{X},\vec{\zeta}))) = \Psi(\vec{X}+\vec{\zeta},\vec{0},\bs{F}(\vec{u}(\vec{X},\vec{\zeta}))) \approx
\Psi(\vec{X} + \vec{\zeta},\vec{0},\bs{F}(\vec{\widetilde{u}}(\vec{X},\vec{\zeta},\vec{0}))).
\label{fe2:eq4d}
\end{equation}
The rightmost form of the energy density in Eq.~\eqref{fe2:eq4d} clearly needs to be evaluated for the reference microstructure only (i.e.~$\vec{\zeta} = \vec{0}$), over a small spatial neighbourhood of a point~$\vec{X}$ (in fact, over the periodic cell~$Q$).

The assumption of periodicity of the microfluctuation component requires that~$\vec{w}$ is periodic, superposed on the approximated~$\vec{v}_0+v_1\vec{\varphi}_1$ field (recall Eq.~\eqref{fe2:eq2}), meaning that
\begin{equation}
\vec{w}(\Gamma_3) = \vec{w}(\Gamma_1), \quad \mbox{and} \quad \vec{w}(\Gamma_2) = \vec{w}(\Gamma_4),
\label{periodicity}
\end{equation}
holds, whereas the four corner points~$P_i$ of the periodic cell~$Q$ are fixed, i.e.~$\vec{w}(P_i) = \vec{0}$, $i = 1, \dots, 4$. In Eq.~\eqref{periodicity}, $\Gamma_i$, $i = 1, \dots, 4$, denote boundary segments of the periodic cell~$Q$, cf. Fig.~\ref{results:fig1c}. In order to alleviate uniqueness concerns (recall Eq.~\eqref{micromorphic:eq22} and the discussion therein), the orthogonality condition~$\langle \vec{\nabla}\vec{w},\vec{\varphi}_1 \rangle_2 = 0$ still needs to be enforced in addition to the periodicity of~$\vec{w}$, although it is now considered in space rather than in translations.
%
%
\subsection{Euler--Lagrange Equations}
\label{approx_energy}
It is important to realize at this point that by the three assumptions made above, the integration over all translations~$\vec{\zeta}$ effectively reduces to a spatial integration over some local neighbourhood in space in one particular realization. To indicate this more clearly, a change of variables~$\vec{X}_\mathrm{m} = \vec{\zeta}$ is employed hereafter. The integration over~$Q$ then reduces to a spatial integration over~$\Omega_\mathrm{m}$, which again spans the~$2\ell \times 2\ell$ periodic cell, and the differentiation~$\vec{\nabla}_\mathrm{m}$ with respect to the local variable~$\Delta\vec{X}$ in Eq.~\eqref{fe2:eq4c} reduces to~$\vec{\nabla}_\mathrm{m} = \partial/\partial X_{\mathrm{m},i}$. For further convenience, and in order to establish a link to computational homogenization, $\Omega_\mathrm{m}$ will be referred to as RVE, which is translated such that its centre point is located at a `macroscopic' position~$\vec{X}$. Furthermore, $\vec{X}_\mathrm{m}$ will be referred to as the `microscopic' coordinate. 

Taking all considerations into account, the approximate average energy takes the form
\begin{equation}
\widetilde{\overline{\mathcal{E}}}(\vec{\widehat{\widetilde{u}}}(\vec{X},\vec{X}_\mathrm{m},\vec{0})) = \frac{1}{|\Omega_\mathrm{m}|}\int_\Omega \int_{\Omega_\mathrm{m}} \Psi(\vec{X}+\vec{X}_\mathrm{m},\vec{0},\bs{F}(\vec{\widehat{\widetilde{u}}}(\vec{X},\vec{X}_\mathrm{m},\vec{0}))) \,\mathrm{d}\vec{X}_\mathrm{m}\mathrm{d}\vec{X},
\label{fe2:eq6}
\end{equation}
in analogy to the exact average energy functional defined in Eq.~\eqref{micromorphic:eq3}. Setting the first variation of~$\widetilde{\overline{\mathcal{E}}}$ equal to zero provides
\begin{equation}
\begin{aligned}
0 &= \delta\widetilde{\overline{\mathcal{E}}}(\vec{v}_0,v_1,\vec{w};\delta\vec{v}_0,\delta v_1,\delta\vec{w}) \\
&=
\left. \frac{\mathrm{d}}{\mathrm{d}h} \widetilde{\overline{\mathcal{E}}}\big(\vec{v}_0+h\delta\vec{v}_0 + \big[v_1+h\delta v_1 + \vec{X}_\mathrm{m}\cdot(\vec{\nabla}v_1 + h\vec{\nabla}\delta v_1)\big]\vec{\varphi}_1+\vec{w}+h\delta\vec{w}\big)\right|_{h = 0} \\
&=
\frac{1}{|\Omega_\mathrm{m}|}\int_\Omega \int_{\Omega_\mathrm{m}} \underbrace{\frac{\partial \Psi(\vec{X}+\vec{X}_\mathrm{m},\vec{0},\bs{F})}{\partial \bs{F}^\mathsf{T}}}_{\bs{P}_\mathrm{m}(\vec{X},\vec{X}_\mathrm{m})}
:
\bigg\{\vec{\nabla}\delta\vec{v}_0(\vec{X})+\vec{\nabla}\delta v_1(\vec{X})\vec{\varphi}_1(\vec{X}+\vec{X}_\mathrm{m},\vec{0})\\
&+
\big[\delta v_1(\vec{X})+\vec{X}_\mathrm{m}\cdot\vec{\nabla}\delta v_1(\vec{X})\big]\vec{\nabla}_\mathrm{m}\vec{\varphi}_1(\vec{X}+\vec{X}_\mathrm{m},\vec{0})+\vec{\nabla}_\mathrm{m}\delta\vec{w}(\vec{X}+\vec{X}_\mathrm{m},\vec{0})\bigg\}\,\mathrm{d}\vec{X}_\mathrm{m}\mathrm{d}\vec{X},
\end{aligned}
\label{fe2:eq7}  
\end{equation}
which differs from the variation of the exact average energy functional in Eqs.~\eqref{micromorphic:eq8} and~\eqref{r_zeta} in only one term, namely in~$\vec{X}_\mathrm{m}\cdot\vec{\nabla}\delta v_1\vec{\nabla}_\mathrm{m}\vec{\varphi}_1$, and in the fact that~$\delta\vec{w}$ is periodic. Following therefore exactly the same procedure as in Section~\ref{sec3_ELeqn}, the set of governing equations for~$\vec{v}_0$ and~$v_1$ is derived, cf. Eq.~\eqref{micromorphic:eq9}, where the definitions of the homogenized stresses now read
\begin{equation}
\begin{aligned}
\bs{P}_1(\vec{X}) &= \frac{1}{|\Omega_\mathrm{m}|}\int_{\Omega_\mathrm{m}} \bs{P}_\mathrm{m}(\vec{X},\vec{X}_\mathrm{m}) \, \mathrm{d}\vec{X}_\mathrm{m}, \quad
P_2(\vec{X}) = \frac{1}{|\Omega_\mathrm{m}|}\int_{\Omega_\mathrm{m}} \bs{P}_\mathrm{m}(\vec{X},\vec{X}_\mathrm{m}):\vec{\nabla}_\mathrm{m}\vec{\varphi}_1(\vec{X}_\mathrm{m},\vec{0}) \, \mathrm{d}\vec{X}_\mathrm{m}, \\
\vec{P}_3(\vec{X}) &= \frac{1}{|\Omega_\mathrm{m}|}\int_{\Omega_\mathrm{m}} \bs{P}_\mathrm{m}^\mathsf{T}(\vec{X},\vec{X}_\mathrm{m})\cdot\vec{\varphi}_1(\vec{X}_\mathrm{m},\vec{0}) + \vec{X}_\mathrm{m}[\bs{P}_\mathrm{m}(\vec{X},\vec{X}_\mathrm{m}):\vec{\nabla}_\mathrm{m}\vec{\varphi}_1(\vec{X}_\mathrm{m},\vec{0})] \, \mathrm{d}\vec{X}_\mathrm{m}.
\end{aligned}
\label{fe2:eq10}  
\end{equation}
In Eq.~\eqref{fe2:eq10}, the stress quantities~$\bs{P}_1$ and~$P_2$ are analogous to their exact counterparts in Eq.~\eqref{micromorphic:eq13}, whereas~$\vec{P}_3$ has an extra term due to the extra contribution~$\vec{X}_\mathrm{m}\cdot\vec{\nabla}\delta v_1\vec{\nabla}_\mathrm{m}\vec{\varphi}_1$ in~$\delta\overline{\mathcal{E}}$, cf. Eq.~\eqref{fe2:eq7}. The spatial dependence of individual quantities in Eq.~\eqref{fe2:eq10} is to be understood as follows. Each macroscopic point~$\vec{X}$ has assigned its own RVE, $\Omega_\mathrm{m}$, over which the local stress~$\bs{P}_\mathrm{m}(\vec{X},\vec{X}_\mathrm{m})$ is defined; hence the dependence of~$\bs{P}_\mathrm{m}$ on both variables~$\vec{X}$ and~$\vec{X}_\mathrm{m}$. On the other hand, the fluctuation field~$\vec{\varphi}_1$ is considered relative to the RVE's centre point, which is assumed to be~$\vec{0}$ for convenience (recall that all RVEs are translated in space such that their centre points coincide with~$\vec{X}$); hence~$\vec{\varphi}_1$ eventually only depends on~$\vec{X}_\mathrm{m}$. The same explanation also holds for~$\vec{w}$. Note that the adopted approach may reflect situations in which individual RVEs (and hence also their fields~$\vec{\varphi}_1$ and~$\vec{w}$), corresponding to a pair of closely positioned macroscopic points, overlap.

At the microscale, the following equation holds
\begin{equation}
\delta\vec{w}:\ 
\vec{\nabla}_\mathrm{m}\cdot\bs{P}_\mathrm{m}^\mathsf{T} = \vec{0}, \quad \mbox{in}\ \Omega_\mathrm{m}, 
\label{fe2:eq12}
\end{equation}
where the orthogonality constraint~$\langle \vec{w},\vec{\varphi}_1\rangle_2 = 0$ and periodicity of~$\vec{w}$ over~$\Omega_\mathrm{m}$ need to be enforced in addition (recall Eqs.~\eqref{micromorphic:eq22} and~\eqref{periodicity}, and the discussions therein). The orthogonality condition is now considered in space over~$\Omega_\mathrm{m}$ instead of in translations~$\vec{\zeta}$, as microstructural translations~$\vec{\zeta}$ have been reduced to changes in spatial position~$\vec{X}_\mathrm{m}$ relative to~$\vec{X}$ in Section~\ref{assumptions}. The periodicity of~$\vec{w}$ and~$\delta\vec{w}$ further entails that the boundary term, similar to the one in Eq.~\eqref{micromorphic:eq10} and now expressed as~$\bs{P}_\mathrm{m}\cdot\vec{N}_\mathrm{m} = \vec{0}$ on~$\Gamma_\mathrm{m}$, disappears. From Eq.~\eqref{fe2:eq12} it is furthermore clear that all microscopic quantities are considered for one realization only.
%
%
\subsection{Computational Homogenization Framework}
\label{computational_framewrok}
\begin{figure}
	\centering
	\def\svgwidth{0.7\textwidth}
	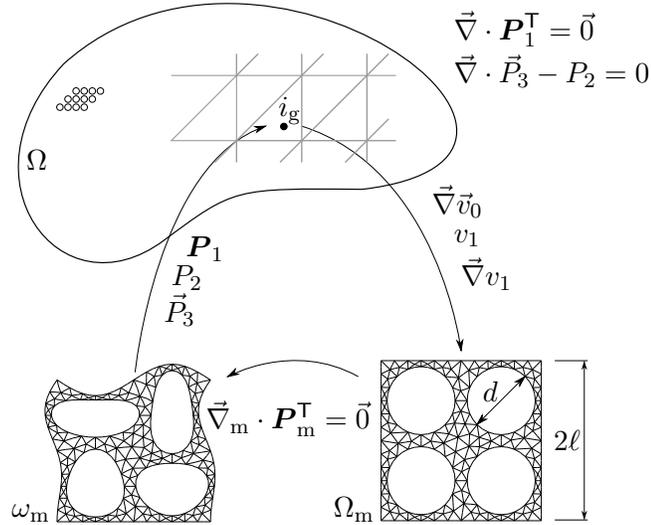
	\caption{A sketch of the micromorphic computational homogenization scheme. Underlying kinematic quantities~$\vec{\nabla}\vec{v}_0$, $v_1$, and~$\vec{\nabla}v_1$, are prescribed to a representative volume element~$\Omega_\mathrm{m}$ considered at each macroscopic Gauss integration point~$i_\mathrm{g}$, where the microfluctuation field~$\vec{w}$ is computed and condensed out. Homogenized properties (i.e.~stresses) are transferred back to the macroscale.}
	\label{fe2:fig1}
\end{figure}

The outline of the solution scheme adopted for the multiscale computational homogenization is given in Algorithm~\ref{fe2:alg1} and the overall procedure is sketched in Fig.~\ref{fe2:fig1}. At each macroscopic Gauss integration point, three quantities are sampled at the macroscale and prescribed to the microscale, namely~$\vec{\nabla}\vec{v}_0$, $v_1$, and~$\vec{\nabla}v_1$ (note that~$\vec{v}_0$ is irrelevant as it does not affect the energy). At the microscale, i.e. inside~$\Omega_\mathrm{m}$, the mode~$\vec{\varphi}_1$ is introduced, and the underlying smooth field constructed. Considering the necessary orthogonality conditions, the microfluctuation field~$\vec{w}$ is computed, which allows for the computation of local stresses~$\bs{P}_\mathrm{m}$. The homogenized stresses are subsequently evaluated according to Eq.~\eqref{fe2:eq10}, and passed on to the macroscale. Here, the global governing equations are assembled and solved.
\begin{algorithm}
  \begin{minipage}{\linewidth}
	\caption{Nested solution scheme for the Computational Homogenization Approach.}
	\label{fe2:alg1}
	\centering
	\vspace{-\topsep}
	\begin{enumerate}[1:]
		\item \textbf{Initialization:}
			\begin{enumerate}[(i):]
				\item Initialize a macroscopic model, $\vec{u}_0(t = 0) = \vec{0}$, $v_1(t=0) = \varepsilon$, $\varepsilon$ a small perturbation.
				\item To each Gauss integration point of the macro-model, assign an RVE.
			\end{enumerate}
		\item \textbf{for~$k=1,\dots,n_T$} (loop over all time steps)
			\begin{enumerate}[(i):]
				\item Apply macroscopic boundary conditions at time step~$k$.
				\item \textbf{while~$\epsilon > \mathrm{TOL}$} (macroscopic solver, iteration~$l$)
				\begin{enumerate}[(a):]
					\item From~$\vec{v}_0^{\,l}$, $v_1^l$  compute for each macroscopic Gauss point~$i_\mathrm{g}$ its deformation gradient~$\bs{I}+(\vec{\nabla}\vec{v}_0^{\,i_\mathrm{g}})^\mathsf{T}$, modal magnitude~$v_1^{i_\mathrm{g}}$, and modal gradient~$\vec{\nabla}v_1^{i_\mathrm{g}}$.
					\item For each macroscopic Gauss point perform RVE analysis:
						\begin{itemize}[\textbf{-}]
							\item Apply underlying deformation dictated by~$\bs{I}+(\vec{\nabla}\vec{v}_0^{\,i_\mathrm{g}})^\mathsf{T}$, $v_1^{i_\mathrm{g}}$, $\vec{\nabla}v_1^{i_\mathrm{g}}$, and~$\vec{\varphi}_1$.
							\item Assemble and solve RVE problem.\footnote{Solution of the RVE problem requires an independent (Newton, or other path-following algorithmic) loop with assembly of the microscopic gradients~$\underline{g}_\mathrm{m}$ (and Hessians~$\bs{\mathsf{H}}_\mathrm{m}$), a result of contributions from all microscopic Gauss integration points. In addition, orthogonality (i.e.~$\langle \vec{w},\vec{\varphi}_1 \rangle_2 = 0$) and periodicity constraints over~$\Omega_\mathrm{m}$ are enforced for~$\vec{w}$.}
							\item Average resulting microscopic quantities to obtain~$\bs{P}_1^{i_\mathrm{g}}$, $P_2^{i_\mathrm{g}}$, and~$\vec{P}_3^{i_\mathrm{g}}$.
						\end{itemize}
					\item Assemble the macroscopic gradient~$\underline{g}^l$ and approximate Hessian~$\bs{\mathsf{B}}^l$ (using, e.g., the BFGS method) from contributions of all macroscopic Gauss points (i.e. from all RVEs).
					\item Update the macroscopic displacements~$\underline{u}^{l+1} = \underline{u}^{l} + \alpha^lp^l $, where the step length~$\alpha^l$ is computed by an inexact line search method along the search direction~$p^l = -(\bs{\mathsf{B}}^l)^{-1} \underline{g}^l$.
					\item Update the iteration error~$\epsilon = \|\underline{g}^l \|_2$.
				\end{enumerate}	
				\item \textbf{end while}
		\end{enumerate}
		\item \textbf{end for}
	\end{enumerate}
	\vspace{-\topsep}
  \end{minipage}
\end{algorithm}

Note that exactly the same set of multiscale governing equations can be obtained by using arguments of computational homogenization, see e.g.~\citep{Kouznetsova:2004}. Starting directly from a two-scale decomposition of the form
\begin{equation}
\vec{u}(\vec{X}_{\mathrm{m}}, \vec{X}) \approx \vec{v}_0(\vec{X}) + \vec{X}_{\mathrm{m}}\cdot\vec{\nabla}\vec{v}_0(\vec{X})
+ \big[v_1(\vec{X}) + \vec{X}_{\mathrm{m}}\cdot\vec{\nabla}v_1(\vec{X})\big]\vec{\varphi}_1(\vec{X}_{\mathrm{m}},\vec{0}) + \vec{w}(\vec{X}_{\mathrm{m}}, \vec{0}),
\end{equation}
where~$\vec{\nabla}\vec v_0$, $v_1$, and~$\vec{\nabla} v_1$ are macroscopic quantities assumed to be constant within each RVE, one can arrive to Eqs.~\eqref{micromorphic:eq9}, \eqref{fe2:eq10}, and~\eqref{fe2:eq12}, by following standard procedures of computational homogenization. We remark, however, that the derivation based on microstructural translations adopted in this paper is (until making approximations on~$\vec{w}$) more rigorous, and that alternative assumptions may lead to implementations which are not comparable with computational homogenization and thus leave room for improvement. Such considerations are, nevertheless, outside the scope of this contribution.
%
%
\section{Results and Comparison with Full Scale Simulations}
\label{results}
\begin{figure}
	\centering
	\subfloat[uniform discretization]{\includegraphics[scale=1]{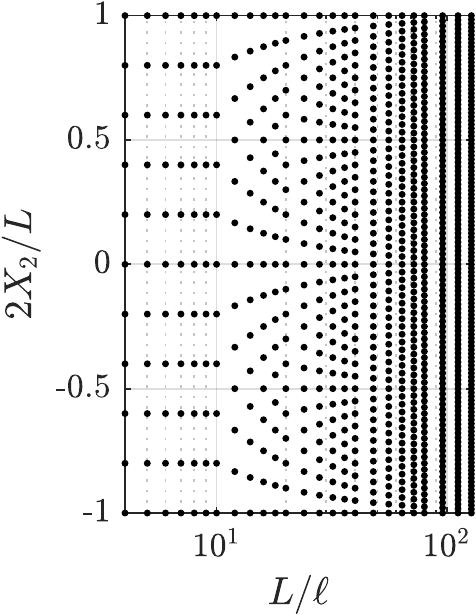}\label{results:fig1a}}\hspace{0.5em}
    \subfloat[coarse discretization]{\includegraphics[scale=1]{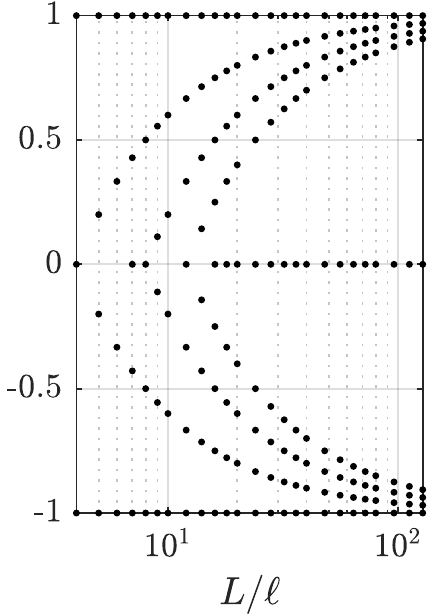}\label{results:fig1b}}
	\subfloat[periodic cell (RVE) discretization]{\def\svgwidth{0.4\textwidth}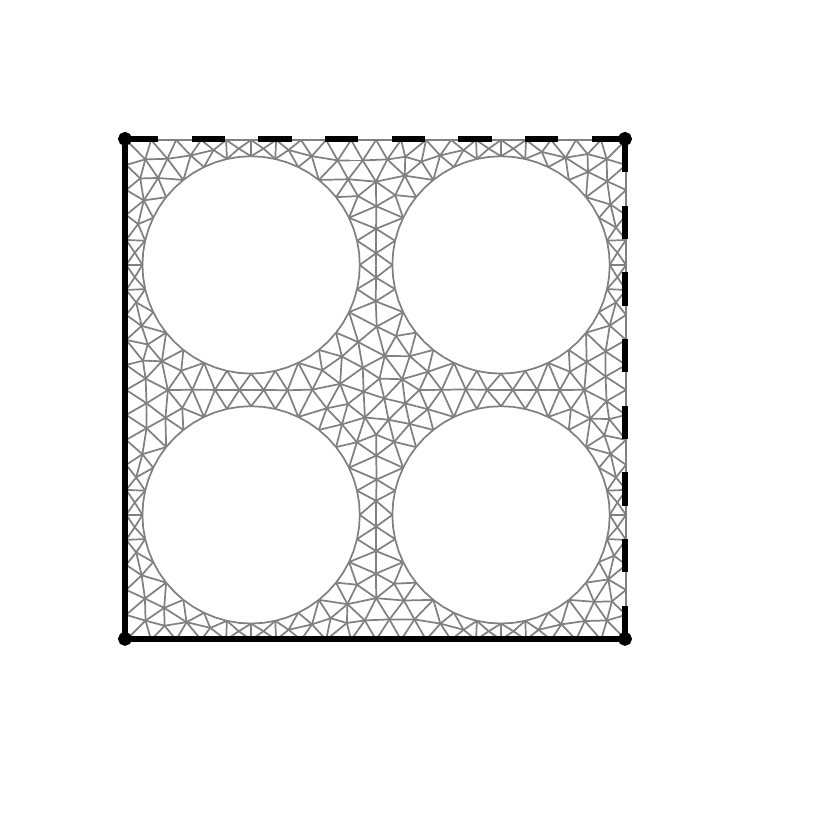\label{results:fig1c}}
	\caption{Macroscopic discretisation (black dots denote individual nodes), uniform~(a) and coarse~(b), along with microscopic discretisation of the RVE domain~$\Omega_\mathrm{m}$ in~(c).}
	\label{results:fig1}
\end{figure}

Following the same procedure as the one outlined in Section~\ref{simplified:discretization}, both coarse quantities~$\vec{v}_0$ and~$v_1$ are discretised by one-dimensional piecewise affine elements in order to implement the computational homogenization approach. Two discretisation schemes along the specimen height are adopted, as sketched in Fig.~\ref{results:fig1}. The first one is uniform, with element size~$\ell$ for scale ratios larger than~$10$, whereas~$10$ elements are used for all scale ratios smaller than~$10$ (cf. Fig.~\ref{results:fig1a}). Such a discretisation is certainly excessively fine, and hence serves as a reference solution. The second discretisation, referred to as coarse, uses prior knowledge on the thickness and location of the two boundary layers, while keeping the size of the smallest macroscopic element to be~$2\ell$ (except for small odd scale ratios), see Fig.~\ref{results:fig1b}. Finally, in Fig.~\ref{results:fig1c}, the RVE discretisation of the~$2\ell \times 2\ell$ periodic cell~$\Omega_\mathrm{m}$ is shown, which makes use of quadratic isoparametric triangular elements. This discretisation has the same typical element size as the one used to obtain the DNS results discussed in Section~\ref{full_problem}. All constraints for~$\vec{w}$, i.e. periodicity in Eq.~\eqref{periodicity} and orthogonality~$\langle \vec{w},\vec{\varphi}_1 \rangle_2 = 0$, are treated as classical equality constraints. All microfluctuation fields~$\vec{w}$ are condensed out for each macroscopic integration point~$i_\mathrm{g}$ by means of an equality constrained Newton-based minimization algorithm, cf. e.g.~\cite{BonnansOptim} or~\cite{Nocedal:Optim}, whereas the macroscopic governing equations are solved using again a quasi-Newton solver.

\begin{figure}
	\centering
	\hspace{2em}\includegraphics[scale=1]{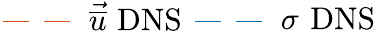}\\
	\hspace{2em}\includegraphics[scale=1]{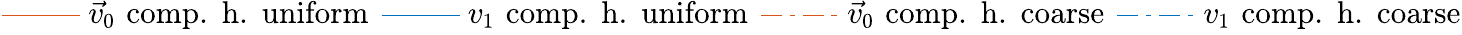}\vspace{-0.75em}\\
    \subfloat[\stackunder{$L/\ell = 5$,}{$u_\mathrm{D}/L = 0.02$}]{\includegraphics[scale=1]{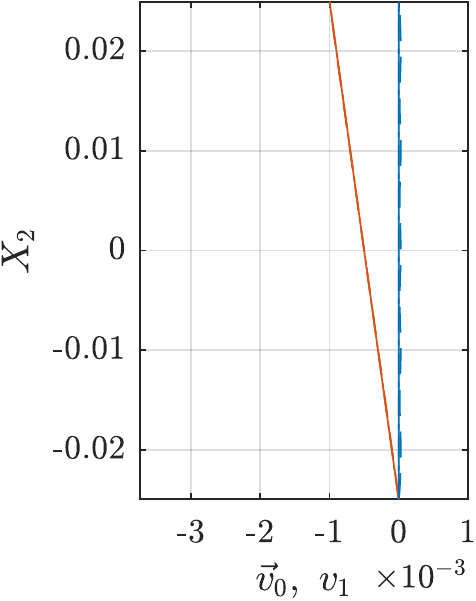}\label{results:fig2b}}	
    \subfloat[\stackunder{$L/\ell = 5$,}{$u_\mathrm{D}/L = 0.075$}]{\includegraphics[scale=1]{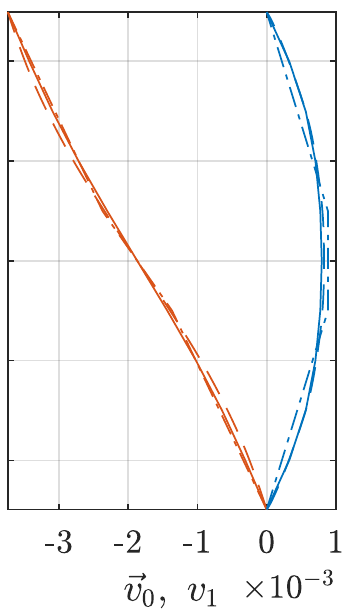}\label{results:fig2a}}
	\subfloat[\stackunder{$L/\ell = 12$,}{$u_\mathrm{D}/L = 0.02$}]{\includegraphics[scale=1]{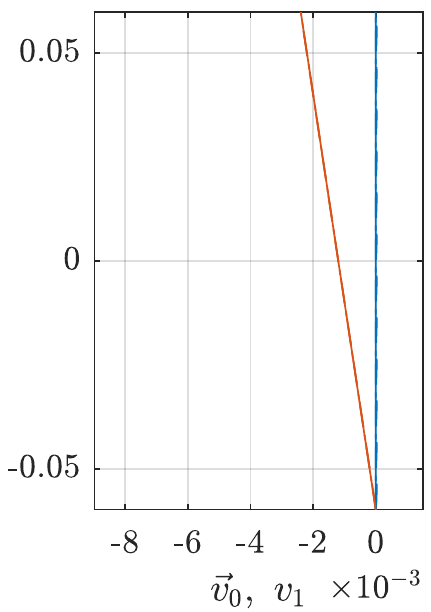}\label{results:fig2d}}
	\subfloat[\stackunder{$L/\ell = 12$,}{$u_\mathrm{D}/L = 0.075$}]{\includegraphics[scale=1]{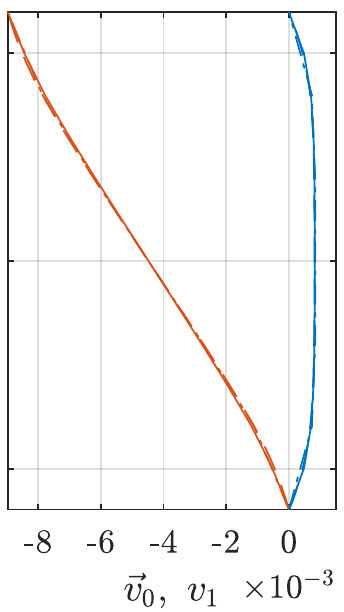}\label{results:fig2c}}
	\caption{The effective fields~$\vec{v}_0$ (vertical component) and~$v_1$ obtained through computational homogenization in comparison with the DNS results for scale ratios~$L/\ell = 5$ in~(a) and~(b), and~$12$ in~(c) and~(d). Two levels of nominal strain, corresponding to the linear ($u_\mathrm{D}/L = 0.02$) and post-bifurcation ($u_\mathrm{D}/L = 0.075$) regime are shown.}
	\label{results:fig2}
\end{figure}

\begin{figure}
	\centering
	\subfloat[$L/\ell = 5$]{\def\svgwidth{0.45\textwidth}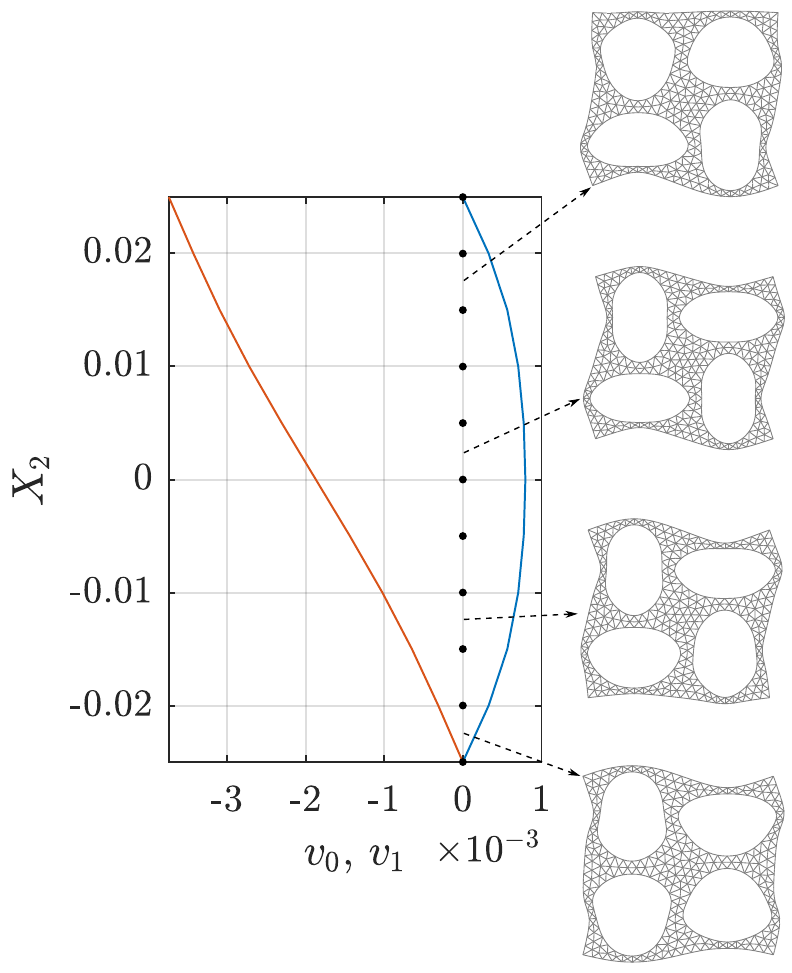\label{results:fig3a}}\hspace{1em}
	\subfloat[$L/\ell = 12$]{\def\svgwidth{0.45\textwidth}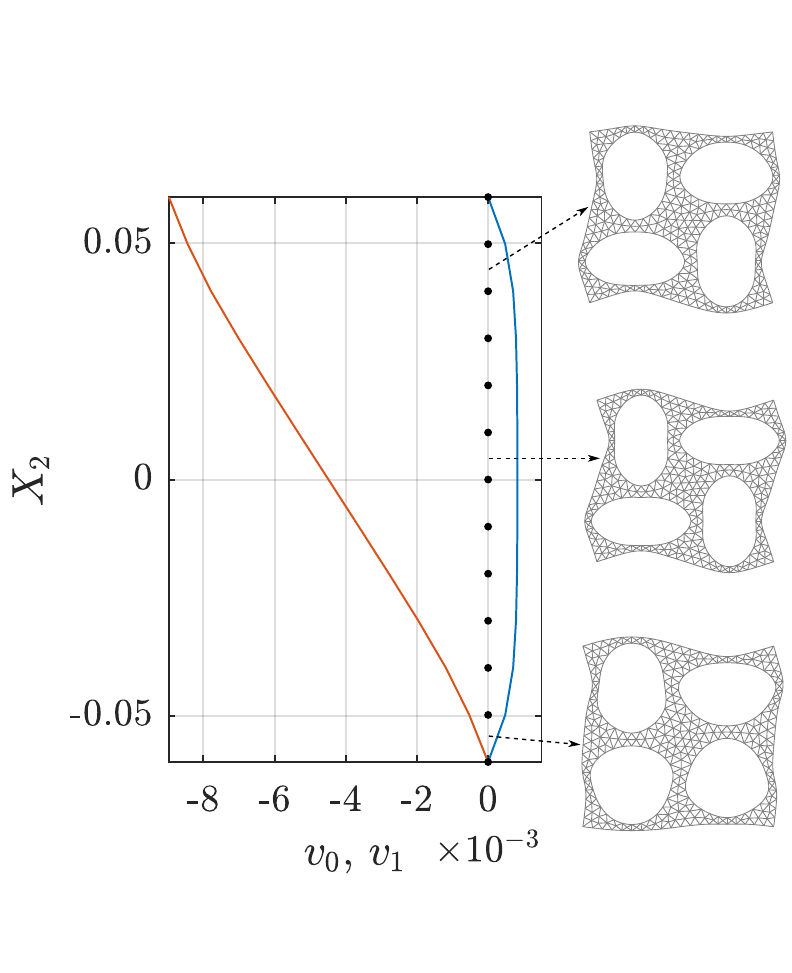\label{results:fig3b}}
	\caption{The effective fields~$\vec{v}_0$ and~$v_1$ at~$u_\mathrm{D}/L = 0.075$, along with corresponding deformed RVEs associated with several Gauss integration points, for scale ratios~$L/\ell = 5$ and~$12$.}
	\label{results:fig3}
\end{figure}

\begin{figure}
	\centering
	\includegraphics[scale=1]{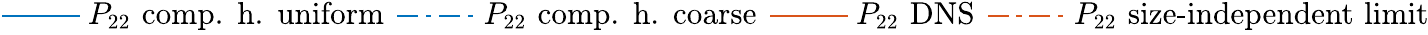}\vspace{-0.5em}\\
    \subfloat[$u_\mathrm{D}/L = 0.02$]{\includegraphics[scale=1]{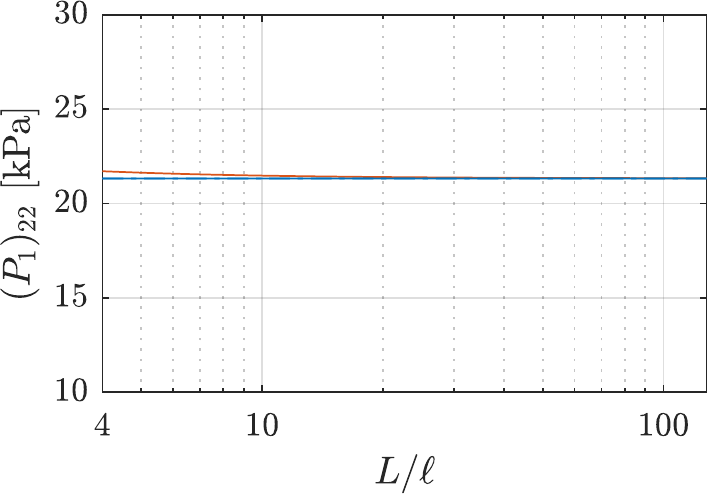}\label{results:fig4b}}\hspace{1em}	
	\subfloat[$u_\mathrm{D}/L = 0.075$]{\includegraphics[scale=1]{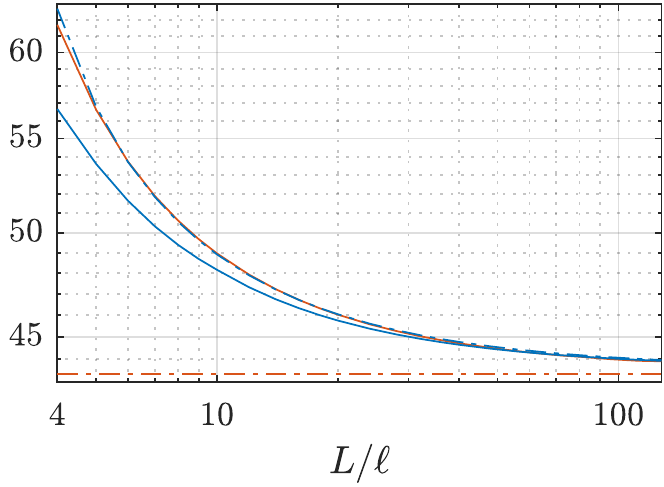}\label{results:fig4a}}
	\caption{Nominal stress~$(P_1)_{22}$ versus scale ratio~$L/\ell$ at two applied strain levels of~(a) $u_\mathrm{D}/L = 0.02$ and~(b) $u_\mathrm{D}/L = 0.075$. Note that in the linear regime, all homogenization approaches coincide (the blue horizontal line).}
	\label{results:fig4}
\end{figure}

Computed results in terms of the coarse quantities~$\vec{v}_0$ and~$v_1$ are compared against the DNS data~$\vec{\overline{u}}$ and~$\sigma$ in Fig.~\ref{results:fig2}. An adequate approximation is now achieved in both the linear and in the post-bifurcation regimes. The boundary layers are captured properly, even for the coarse discretisation. In the linear regime, $\sigma \neq 0$ again as a consequence of the discrepancy in its definition compared to that of~$v_1$. The homogenized field~$v_1$ does vanish here as expected, but unlike in the closed-form homogenized solutions of Section~\ref{simplified}. The computed responses of the two uniform discretisations corresponding to scale ratios~$L/\ell = 5$ and~$12$ and an applied strain of~$u_\mathrm{D}/L = 0.075$ are further shown with the corresponding deformed RVEs in a couple of integration points in Fig.~\ref{results:fig3}. At the centre, where~$v_1$ is constant, the cells deform practically periodically, according to the pattern given by~$\vec{\varphi}_1$. But closer to the boundaries~$v_1$ approaches zero and hence the pattern becomes less prominent. Since~$v_1$ varies significantly on the size of the RVE, $2\ell$, the RVE deforms non-periodically. 

Finally, the homogenized stresses are reported in Fig.~\ref{results:fig4}, where a significant improvement compared to the closed-form homogenization approach presented in Section~\ref{simplified} can be observed. In the linear regime, the modulation function~$v_1 = 0$, whereas~$\vec{w}$ is periodic on top of an underlying affine deformation~$\vec{v}_0$. This effectively results in an FE\textsuperscript{2} type of approximation, with a relative error below~$2\,\%$, and practically no size effect, cf. Fig.~\ref{results:fig4b}. In the post-bifurcation regime, $v_1 \neq 0$, and individual RVEs buckle in a synchronized manner coordinated through the term~$v_1\vec{\varphi}_1$. The microfluctuation field~$\vec{w}$ still compensates any local fluctuations on top of the approximate quantity~$\vec{v}_0+v_1\vec{\varphi}_1$, significantly relaxing the system. This results in nominal stresses~$(P_1)_{22}$ that are within~$10\,\%$ of relative error in the case of the regular macroscopic discretisation, which is systematically more compliant compared to the DNS results. Note further that the largest error occurs for the smallest scale ratio considered, i.e.~$L/\ell = 4$. The coarse discretisation is, on the other hand, somewhat stiffer, and provides in this particular case a slightly more accurate result, cf. Fig.~\ref{results:fig4a}.
%
%
\section{Summary and Conclusions}
\label{summary}
This paper presented a micromorphic homogenization framework, capable of capturing long-range correlated fluctuation fields emerging in elastomeric mechanical metamaterials due to local buckling of their microstructure. The ensemble averaging based homogenization framework, the emergent Euler--Lagrange equations and the numerical approximation schemes have been elaborated and commented. Results have been compared against averaged full-scale numerical simulations reported by~\cite{Ameen2018}.

The key developments and results can be summarized as follows:
\begin{enumerate}
	\item A kinematic decomposition of the underlying displacement field considered for an entire family of translated microstructures has been proposed.
	
	\item The proposed pattern-based decomposition, upon making use of the variational homogenization framework and ensemble averaging, has provided an effective micromorphic continuum enriched with the magnitude of the underlying long-range correlated fluctuation field. The additional micromorphic field establishes kinematic interactions between individual cells, which is essential for accurate homogenization and capturing quantitative size effects.
	
	\item A computational homogenization scheme has been developed, making use of assumptions on the local energy density, smoothness properties of the effective kinematic fields, and the periodicity of the microfluctuation field. This allowed for an efficient and accurate solution of the macroscopic governing equations.
	
	\item The key aspects of the implementation of the proposed framework have been discussed, focusing in particular on the effective fields~$\vec{v}_0$ and~$v_1$, along with the appropriate boundary conditions and orthogonality constraints.
	
	\item Using a representative example, the ability of the proposed methodology to capture accurate solutions has been demonstrated in terms of kinematics as well as stress quantities. Homogenized stresses were always within~$10\,\%$ relative error compared to the direct numerical simulations, capturing relevant size effects.
\end{enumerate}
The proposed methodology, although presented on a specific case, can easily be extended to more complex systems, containing multiple long-range correlated fluctuation fields and exhibiting complex size effects.
%
%
%
%
\section*{Acknowledgements}
The research leading to these results has received funding from the H2020 European Research Council under the European Union's Seventh Framework Programme (FP7/2007-2013)/ERC grant agreement \textnumero~[339392].
%
%

%
\end{document}